\documentclass{aastex}

\usepackage{emulateapj5,apjfonts}
\usepackage{graphicx}
\usepackage{onecolfloat}

\begin{document}

\def\gsim { \lower .75ex \hbox{$\sim$} \llap{\raise .27ex \hbox{$>$}} }
\def\lsim { \lower .75ex \hbox{$\sim$} \llap{\raise .27ex \hbox{$<$}} }

\slugcomment{accepted for publication in The Astrophysical Journal}
\shorttitle{Crowded Field 3D Spectrophotometry of XPNe in M31}
\shortauthors{Roth et al.}

\twocolumn[ 

\title{3D Spectrophotometry of Planetary Nebulae in the Bulge of M31}

\author{Martin M. Roth\altaffilmark{1,2}, Thomas Becker\altaffilmark{1,2}, 
        Andreas Kelz \altaffilmark{1}}
\affil{Astrophysikalisches Institut Potsdam, An der Sternwarte 16, D-14482 Potsdam, Germany}
\email{mmroth@aip.de}
\and
\author{J\"urgen Schmoll\altaffilmark{1,2,3}}
\affil{Astronomical Instrument Group, Dept. of Physics, University of Durham, \\
Rochester Buildg., South Road, Durham DH1 3LE, United Kingdom}
\email{Jurgen.Schmoll@durham.ac.uk}

\begin{abstract}

   We introduce crowded field integral field (3D) spectrophotometry as a useful
   technique for the study of resolved stellar populations in nearby galaxies.
   The spectroscopy of individual extragalactic stars, which is now feasible 
   with efficient instruments and large telescopes, is confronted with the 
   observational challenge of accurately subtracting the bright, spatially and 
   wavelength-dependent non-uniform background of the underlying galaxy. As a 
   methodological test, we present a pilot study with selected extragalactic 
   planetary nebulae (XPN) in the bulge of M31, demonstrating how 3D spectroscopy is 
   able to improve the limited accuracy of background subtraction which one would 
   normally obtain with classical slit spectroscopy. It is shown that due to
   the absence of slit effects, 3D is a most suitable technique for spectrophometry.
   We present spectra and line intensities for 5 XPN in M31, obtained  with the MPFS 
   instrument at the Russian 6m BTA, INTEGRAL at the WHT, and with PMAS at the
   Calar Alto 3.5m Telescope. 
   The results for two of our targets, for which data are available in the literature,
   are compared with previously published emission line intensities.
   The three remaining PN have been observed spectroscopically for the first
   time. One object is shown to be a previously misidentified supernova remnant. 
   Our monochromatic H$_\alpha$ maps are compared with direct Fabry-P\'erot and 
   narrowband filter images of the bulge of M31, verifying the presence of filamentary 
   emission of the interstellar medium in the vicinity of our objects. We present an 
   example of a flux calibrated and continuum-subtracted filament spectrum and 
   demonstrate how the ISM component introduces systematic errors in the measurement 
   of faint diagnostic PN emission lines when conventional observing techniques are 
   employed. It is shown how these errors can be eliminated with 3D spectroscopy, 
   using the full 2-dimensional spatial information and point spread function (PSF) 
   fitting techniques. Using 3D spectra of bright standard stars, we
   demonstrate that the PSF is sampled with high accuracy, providing a centroiding 
   precision at the milli-arcsec level. Crowded field 3D spectrophotometry and the 
   use of PSF fitting techniques is suggested as the method of choice for a number 
   of similar observational problems, including luminous stars in nearby galaxies, 
   supernovae, QSO host galaxies, gravitationally lensed QSOs, and others.
   
\end{abstract}


 ]

\altaffiltext{1}{Visiting Astronomer, German-Spanish Astronomical Centre, Calar Alto, operated by the Max-Planck-Institute for Astronomy, Heidelberg,
jointly with the Spanish National Commission for Astronomy.}
\altaffiltext{2}{Visiting Astronomer, Special Astrophysical Observatory, Selentchuk,
Russia.}
\altaffiltext{3}{Visiting Astronomer, Isaac Newton Group of Telescopes, La Palma.}

\section{Introduction}
\label{INTRODUCTION}

Understanding the formation and evolution of galaxies is one of the most prominent
goals of modern cosmology. After a long-lasting discussion about the best 
evolutionary scenarios for elliptical galaxies, for example, either according to 
the classical picture of instantaneous dissipative collapse (Larson 1974), or 
rather in the hierarchical scenario through gradual mergers of spirals (White 1980), 
it seems today that the latter view is the more widely accepted one. 
The observational evidence, that nearby ellipticals show peculiarities like 
counter-rotating cores, rings, gas disks, and recent star formation, can be 
explained by merger and accretion processes over an extended period of time. 
On the theoretical side, numerical N-body simulations combined with the SPH technique 
in a $\Lambda$CDM cosmology are capable of reproducing observed photometric and 
kinematical properties of elliptical galaxies fairly well, thus supporting the 
hierarchical picture, despite some remaining discrepancies (e.g. Meza et al.\ 2003).
On the other hand, Chiosi \& Carraro 2002 presented SPH simulations producing
elliptical galaxies, which are compatible with monolithic collapse,
so that the problem is not yet finally solved.

Among the classical tests for the validity of any of the theoretical pictures, 
the integrated-light measurement of broad-band colors and of absorption 
line indices play an important role. The observations can be compared with 
integrated light stellar population models in order to distinguish between 
different star formation histories and chemical enrichment (e.g. Saglia et 
al.\ 2000, Mehlert et al.\ 2000). 

However, the presence of dust and ionized gas in the interstellar medium of
elliptical galaxies (Goudfrooij et al.\ 1994, Bregman et al.\ 1992) has
motivated the question, whether or to which extent the observed 
gradients are real (dust effecting the surface photometry through extinction, 
gaseous emission filling in the absorption line profiles of H$_\beta$, 
Mg{\it b}, Mg$_2$). Moreover, systematic errors of long-slit spectroscopy 
due to the effects of a variable point-spread-function of the spectrograph 
optics along the slit have to be taken into account. For a discussion of 
these problems, see Mehlert et al.\ 2000.

From a different point of view, Freeman \& Bland-Hawthorn (2002) advocate
in what they termed ``near-field cosmology'' to probe the 6-dimensional phase
space and chemical compositions of very many individual stars in order to unravel 
details of the star formation history of the Milky Way and thus the signatures 
of its formation and evolution.

An alternative approach, now rapidly producing new abundance data and reaching out 
to significantly larger distances, is based on the spectroscopic analysis of 
individual resolved luminous stars in nearby galaxies, e.g.\ M31 (Smartt et al.\ 2001, 
Venn et al.\ 2000), NGC6822 (Venn et al.\ 2001), M33 (Monteverde et al.\ 1997), or 
NGC300 (Urbaneja et al.\ 2003). 
Using the ESO-VLT + FORS, the feasibility of high signal-to-noise stellar 
spectroscopy even beyond the local group was demonstrated by Bresolin et 
al.\ (2001), who measured 7 supergiants of spectral types B, A, and F with 
V$\approx$20.5 in NGC3621, a galaxy which is located at a distance of 6.7 Mpc.
For a review on extragalactic stellar spectroscopy, see Kudritzki 1998.

Currently the only way to measure individual abundances from old or intermediate age
stars in galaxies more distant than the Magellanic Clouds is through the emission line 
spectra of extragalactic planetary nebulae (XPN), see Walsh et al.\ 2000.  This approach 
has some similarities with the standard method of measuring abundance gradients from
individual H$\;$II regions in the disk of spiral galaxies (Shaver et al.\ 1983, 
Zaritsky et al.\ 1994). As opposed to H$\;$II regions, 
XPN metallicities can be derived in a homogeneous way for galaxies of {\em any} Hubble 
type, and on all scales of galactocentric distances. The task of obtaining abundance
gradients out to large radii, where a low surface-brightness precludes to 
measure reliable colors or absorption line indices, can be addressed with XPN, 
providing important constraints for galactic evolution models (Worthey 1999).
Moreover, since radial velocities of XPN are measurable out to several effective
radii, they are  potentially useful for probing the gravitational potential 
of galaxies (M{\' e}ndez et al. 2001), and  for tracing merger events (Hui et al. 1995,
Durrell et al.\ 2003).
Recently, XPN and an H$\;$II region have been detected in the intracluster space 
of the Virgo cluster, giving an excellent opportunity to study the properties
of this unique stellar population and, potentially, their star 
formation history and metallicity (Arnaboldi  et al.\ 2002, Gerhard et al.\ 2002, 
Feldmeier et al.\ 2003).

When compared to integrated light studies, the analysis of single objects has 
the advantage that the abundance determinations are based on detailed physical 
models, and that an assessment of individual errors is possible. The study
of resolved stellar populations in galaxies  out to the distance of the Virgo 
cluster has become a major science case for the proposed new generation of 
Extremely Large Telescopes (Wyse et al. 2000, Hawarden et al. 2003, 
Najita \& Strom 2002).



 \begin{figure*}[htb]
 \centering\includegraphics[width=0.70\linewidth,clip]{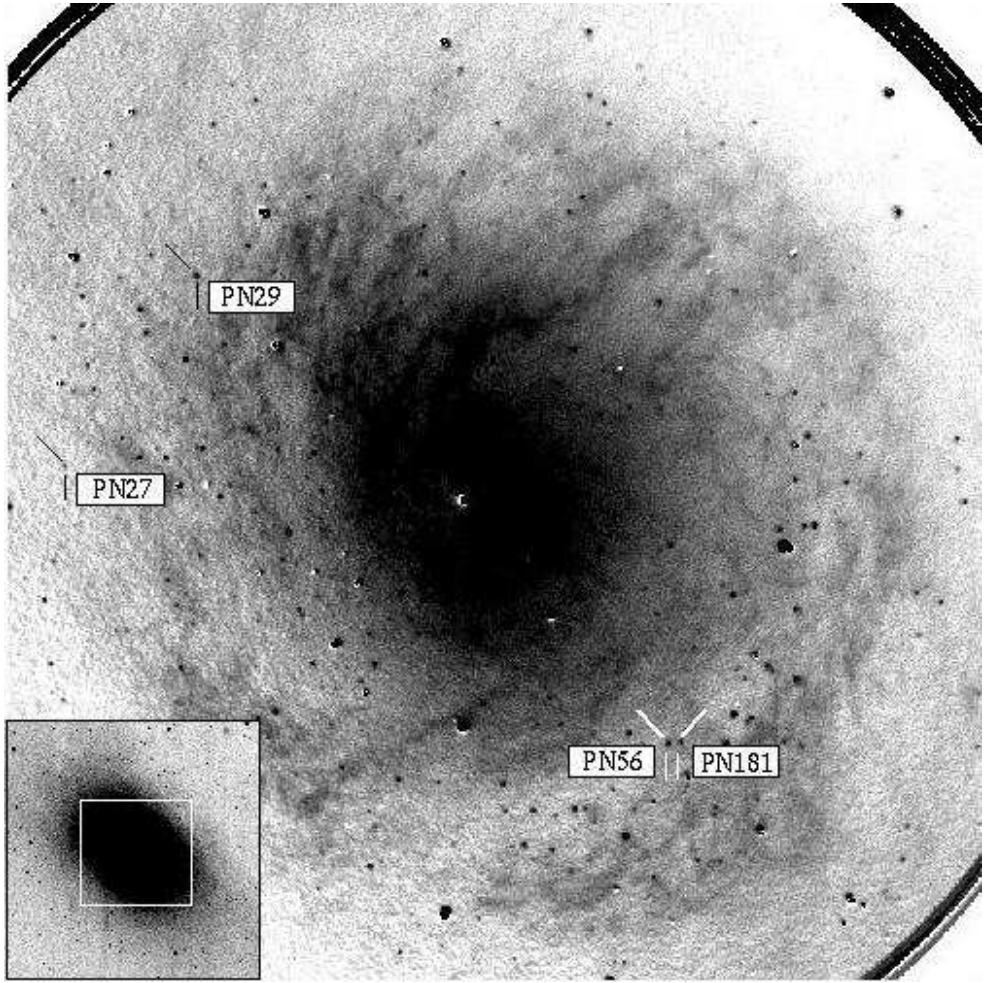}
 \figcaption{H$_\alpha$ narrow-band image of the central 6~arcmin of the 
 Andromeda galaxy M31, obtained with Calar Alto 3.5m prime focus camera + 
 Fabry-P\'erot. The continuum has been subtracted using an off-band image
 to reveal the complex and fuzzy emission structure of the ISM. The tiny dark
 spots are XPN candidates, while spots with white substructures are indicative
 of residuals from imperfectly subtracted continuum sources (stars). The insert 
 shows a broad-band DSS image with an outline of the H$_\alpha$ frame, illustrating
 the rapid increase of continuum surface brightness towards the nucleus. 
 Orientation: North to the top, East to the left.
 \label{M31Halpha}}
 \end{figure*}

However, the spectroscopy of individual point sources in nearby galaxies suffers
generally from source confusion, either from the continuum light of unresolved 
stars with a small angular separation from the target, or from the emission
line spectra of H$\;$II regions and diffuse nebulosities of the interstellar
medium (ISM), or from both components at the same time. 
A common feature of these two background components is that their spectral and
spatial distribution is uncorrelated and inhomogeneous, making it difficult 
to provide accurate background estimates when one is using conventional slit
spectrographs.

Roth et al.\ (1997) proposed the development of 
{\em crowded field integral field (3D) spectroscopy} 
as a superior method to disentangle the spectrum of a target point 
source from the background in a crowded field environment. 3D spectroscopy
(for short: 3D) is an established, but not yet entirely common and therefore
relatively new observing technique. It provides simultaneously spectra for 
many spatial elements over a 2-dimensional field-of-view (FOV), 
see review by Boulesteix (2002). 3D has been used very successfully for kinematic
mapping and integrated light stellar population studies of galaxies 
(de Zeeuw et al.\ 2002).
Our reason for observing {\em point sources} may not be immediately obvious, 
given the technical effort to generate and record up to thousands of spectra in 
any single exposure. The main argument, however, is that the knowledge of the 
point-spread-function (PSF) of a stellar object can be used to apply 
PSF-fitting techniques, thus discriminating the source against the background. 
Such a technique would be analogous to PSF-fitting CCD photometry, which has 
been so successful for the construction of globular clusters CMDs and the 
photometric study of resolved stellar populations in nearby galaxies (Mateo 1998).

To this end, the design of the new integral field spectrophotometer PMAS
was put forward (Roth et al. 2000a), featuring a relatively small FOV 
(max. 32$\times$32 element lenslet array), but a unique wavelength coverage 
(4096 pixels) and a good response down to the UV, compared to other 3D 
instruments (Boulesteix 2002). First PMAS observations of point sources in M31,
which are relevant for the problems discussed below, were presented by Roth 
et al.\ 2002c.

In this paper, we discuss a pilot study for the novel observing technique 
of {\em crowded field 3D spectrophotometry} for the observation of XPN in the 
local group galaxy M31. 
At a distance of 770~kpc towards M31 (Madore \& Freedman 1991), an angular 
separation of 1~arcsec corresponds to a projected linear scale of 3.7~pc, 
making each XPN appear 
as a point source for ground-based observations. In M31, more than 400 XPN
candidates have been detected with on-band/off-band imaging techniques in the 
prominent [O$\;$III] $\lambda$5007 emission line 
(Ciardullo et al.\ 1989, henceforth CJFN89), enabling us to readily
select targets from a large range of apparent brightness 
(m$_{5007} \approx 20.5 \ldots 24.5$) and location over the host galaxy, 
facilitating telescope pointing and identification, and allowing us to 
investigate in detail faint diagnostic lines, which tend to suffer from poor 
background subtraction accuracy. In a future paper (Fabrika et al., in 
preparation) we shall present a similar study for 3D observations of 
luminous stars in M33.

The paper is organized as follows: in \S~\ref{OBSPROBL}, we review the 
peculiar observational problems for the spectroscopy of XPN and explain,
how 3D spectrophotometry can be used to overcome these problems.
\S~\ref{INSTRUMENT} presents a brief overview of 3D techniques and the 
instruments that we have used for this work. We describe our observations
in \S~\ref{OBSERV}, and data reduction/analysis in 
\S\S~\ref{DATARED},\ref{RESULTS}. We discuss our results in \S~\ref{DISCUSSION},
and conclude with a summary in \S~\ref{SUMMARY}.

\section{Observational problems}
\label{OBSPROBL}

\subsection{Previous Work}
\label{PREVIOUS}
XPN are ideal tracers of intermediate age and old extragalactic stellar populations, 
because their hot central stars are among the most 
luminous stars in the HRD, emitting their radiation predominantly 
in the UV. A substantial fraction (of order 10\%) of the total luminosity is 
re-emitted by the surrounding nebula in a prominent emission line spectrum,
which gives enough contrast (for the bright lines) to detect the object as
a point source against the bright background of unresolved stars of the parent
galaxy. A practical application of 
this property consisted in narrow-band imaging spectrophoto\-metry, centered
on the bright emission line  of [O$\,$III] $\lambda$5007, and the construction
of \mbox{PN} luminosity functions (PNLF) for the purpose of distance
determinations (see review by Ciardullo 2003). 
Approximately 5000~XPN in more than 40 galaxies have been identified to
date (Ford et al.\ 2002).

Prompted by the motivation described in \S~\ref{INTRODUCTION}, several authors 
have conducted spectroscopic observations of individual XPN in nearby galaxies 
and derived abundances from the observed emission line intensities. Jacoby \& 
Ciardullo (1999) obtained spectra of 15 XPN in M31 using the Kitt Peak 
4m telescope, equipped with the RC spectrograph and a multi-slit mask, with 
total exposure times of 6.5 and 6~hours, for two setups, respectively.
Their paper gives an excellent account of the technical and observational
issues to be addressed below, and we will use it as our major reference
(henceforth JC99).
Richer et al.\ 1999 (RSM99) observed 30 XPN in M31, and 9 XPN in M32.
Walsh et al.\ 1999 investigated 5 objects in NGC$\;$5128 (the Cen-A galaxy). 
In the best cases, and as far as the data quality allowed, photoionization
models were computed and matched to the observed spectra. From the
constraints imposed on these models, abundances, in particular [O/H],
were derived. A wealth of data exists for Magellanic Cloud objects which
are an order of magnitude closer and therefore much easier to observe than 
those in M31 and other more distant galaxies. 
The LMC study of Dopita et al. \cite{dopita97} 
is an example for the potential of XPN to investigate the chemical evolution 
of stellar populations.
More recently, observations of XPN in M33, obtained with the multi-object
fiber spectrograph at the WHT, were presented by Magrini et al. (2003).
We shall compare our new 3D results with these previous publications in 
\S~\ref{DISCUSSION}.

Although we have pointed out that central stars of XPN are intrinsically
luminous, and that the prominent [O$\;$III] $\lambda$5007 line reflecting this
brightness is relatively easy to detect, the comparison of XPN spectra
with ionization models and subsequent abundance determinations requires
the detection of much fainter lines, e.g.\ [O$\;$III] $\lambda$4363, which 
is important for measuring the electron temperature T$_e$.
For a $\lambda$4363 line intensity of e.g.\ 1\% of $\lambda$5007 
(depending on T$_e$), spectroscopy with a 3.5m telescope will yield a total
rate of $\approx$0.25~photons/sec for an XPN of m$_{5007}$=21.0. At a distance 
modulus of 24.6, this corresponds to an object near the bright 
cut-off of the PNLF. In the case of M31, if this object is less than 2~arcmin
away from the nucleus, the surface brightness is $\mu_V \le 19 $
(Hoessel \& Melnick 1980), contributing typically more than 0.7~photons/sec
within an aperture of 1~arcsec diameter and a spectral resolution of 3~{\AA}.
Obviously, at this level of emission line intensity, the measurements are 
background limited and require exposure times of many hours to determine
the spatially variable surface brightness distribution of unresolved stars
accurately enough. In most real cases, the [O$\;$III] $\lambda$4363 is fainter
and more difficult to measure than in our simple example.
Jacoby \& Kaler \cite{jacobykaler93} (JK93) have described the effects of 
crowding for their observations of XPN in the Magellanic Clouds and summarized 
the following observational problems.

\subsubsection{Slit Losses}
\label{SLITLOSSES}
JK93 present an instructive diagram (their Fig.~1), allowing to immediately 
read-off the percentage of light loss as a function of slit width relative to
the seeing FWHM, depending on centering errors of the star with respect to the 
slit. A slit width identical to the seeing FWHM will in the ideal 
case result in slit losses of roughly 25\%, increasing to 30\%, 50\%, and 70\% 
for a slit offset of 0.25, 0.50, and 0.75 (in units of seeing disk FWHM), 
respectively. Opening the slit will reduce slit losses, allowing for guiding
errors and less precise pointing. However, the spectral resolution will be 
degraded, and again the amount of background contamination will increase. For 
a given background surface brightness it is possible to determine a slit width 
with an optimized signal-to-noise ratio of the detected flux. 
Since total exposure times are spanning many hours, and because of the
variable  nature of seeing (Sarazin \& Tokovinin 2002), it would be 
impractical to constantly adjust the slit width to the optimum. 
In addition, a slit width optimized at a certain wavelength may not be
optimal at another wavelength, e.g.\ when there are strong absorption or 
emission line features in the background.

\subsubsection{Seeing}
\label{SEEING}
Spectrophotometric measurements covering a large spectral range, 
e.g.\ from [O$\,$II] $\lambda$3727
to [S$\,$II] $\lambda$6731, are affected by an increasing size of the seeing disk
from the red to the blue (Fried 1966). In order to avoid substantial slit losses 
in the blue, and, resulting from this effect, systematic wavelength-dependent flux
errors, JK93 advise a combination of exposures with wide {\em and} narrow slit
width settings, thus providing a calibration for these errors while retaining
the required spectral resolution and reduced background contamination for
the measurement of faint lines. In our application, the overall efficiency 
would drop prohibitively low due to the increase of exposure times.

\subsubsection{Atmospheric Refraction}
\label{REFRACTION}
The long exposure times which are generally required for faint XPN
result in observations over a large range of airmass, i.e.\
exposures at large zenith distance cannot be avoided. Atmospheric dispersion 
produces an elongated chromatic focal plane image, whose length is a function 
of zenith distance, leading to systematic errors if some light is not 
collected in the slit.
In multi-slit spectrographs the problem cannot be circumvented with the 
standard strategy of adjusting the slit orientation along the parallactic
angle when the slit masks have a mechanical fixed orientation (JC99).  
The alternative to open the slit larger than what would be necessary on 
the basis of the seeing disk,
automatically results in a penalty of increased background contamination
(see below). An instrumental concept to counteract this problem is the 
atmospheric dispersion compensator (ADC, e.g.\ Avila et al.\ 1997), 
which, however, introduces light-losses and is not always applicable.


\subsubsection{Background Contamination}
\label{BACKGROUND}
The presence of a bright, both spatially and wavelength-dependent variable
background of the host galaxy is a most serious complication for accurate
emission line spectrophotometry of XPN. The following considerations are 
not restricted to these objects. They also hold true in analogous ways 
for stars or any other point sources in galaxies, e.g.\ blue supergiants,
novae, supernovae, H\,II regions, and so forth.

The galaxy background surface brightness distribution is composed of 
unresolved stars and contributions from the interstellar medium (ISM), 
either in the form of dust (extinction), or as gaseous emission, 
which may exhibit both diffuse or filamentary characteristics.
Jacoby et al.\ 1985, for example, have mapped in H$_\alpha$ 
a prominent spiral-shaped gaseous component in the bulge of M31, using a 
mosaic of CCD images. Long-slit spectroscopy described by Ciardullo et al.\ 1988 
reveales strong emission in [S$\,$II] $\lambda\lambda$6717, 6731, and 
[N$\,$II] $\lambda\lambda$6548, 6583 --
unfavourably coinciding with diagnostic emission lines of XPN.

The accuracy which can be obtained from an interpolation of the background
along the slit on both sides next to the target appears to be
limited. JK93 present an example (LMC J22, their Fig.~2 and Fig.~3)
which demonstrates the severe effects arising from a bright nearby star,
and from the ubiquitious diffuse ISM emission. Similar problems have
obviously also affected the extraction of the spectrum of PN5601 in NGC5128
of Walsh et al.\ 1999, as one can deduce from the skewed appearance of the 
continuum (their Fig.~2 and Fig.~3).
Some of the spectra plotted in RSM99 and JC99 show systematic deviations 
of the average continuum from zero, indicative of errors in the 
background subtraction (e.g. RSM99: PN6, PN24, PN25 in M32, PN12, PN23,
PN28 in M31; JC99: CJFN23, 125, 177, 179, 470, FJCHP51 in M31). While
this is not necessarily evidence for errors in the emission line intensities, 
one cannot rule out the possibility that e.g.\ H$_\beta$ or one of the
other Balmer lines could have been affected in the sense that the 
absorption line present in the background was not properly taken into account,
leading to an underestimate of the intensity of this emission line.
We will show that the discrepancy in the [O\,III]/H$_\beta$ line ratio
for PN29 in M31 reported by JC99 between their result and the one of
RSM99 can be explained in this way. Such systematic errors would be  
particularily troublesome since an estimate of the interstellar reddening 
is affected by the accuracy of the Balmer decrement measurement.

Another severe problem is related to gaseous emission of the ISM.
Let us consider, for example, a filament extending perpendicular to 
the slit, and stretching across the point source under study.
This feature will escape completely undetected and will cause  inevitably 
a systematic increase of some diagnostic line intensities of the 
target. Other, less extreme situations may occur when again
the true flux contribution of the filament cannot be estimated
precisely enough because only two measurements next to the object
along the slit are available for an interpolation. Again, PN29  in the
study of JC99 is an example, where an overestimate of the flux
and an unusual intensity ratio of [S$\,$II] $\lambda\lambda$6717,6731 
has led to unphysical predictions for the electron density in this object. 
This case is discussed in more detail in Section \ref{RESULTS}.
Other examples in the sample of JC99 are the faint XPN CJFN470,
and CJFN455, where the [S$\,$II], [O$\,$II] and
[N$\,$II] line intensities are probably overestimates.

\begin{figure}[t!]
\plotone{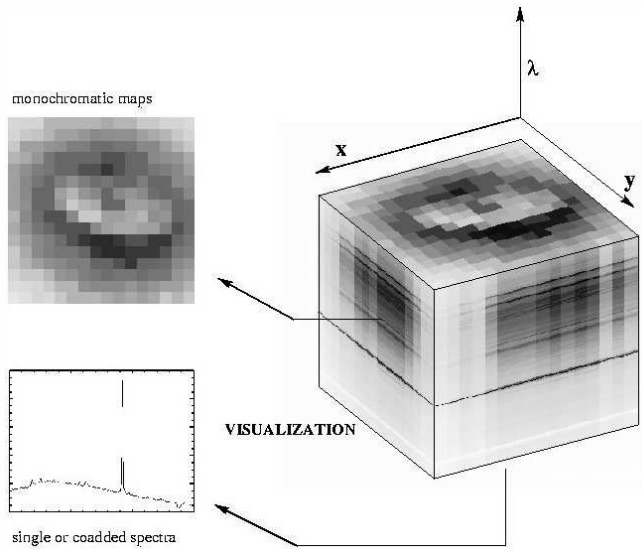}
\caption{Datacube, generated with integral field spectroscopy.  \label{DATACUBE}}
\end{figure}

\subsection{Advances of Integral Field Spectroscopy}

3D spectroscopy, contrary to conventional slit spectroscopy, produces datacubes
with full 2-di\-men\-sio\-nal spatial information within the FOV of the instrument.
Fig.~\ref{DATACUBE} shows an idealized datacube with a square field of view,
square spatial elements (``spaxels'') along coordinates (x,y), and spectra
arranged in columns ($\lambda$). From this entity, one can retrieve monochromatic
maps at certain wavelengths (datacube slices), or co-added maps equivalent to 
broad- or narrow-band images, and single spectra (columns), as well as co-added
spectra for any desired region, i.e.\ digital aperture within the FOV.
We stress that with proper spatial sampling, it is possible to record
intensity profiles of monochromatic maps. For point sources, this is equivalent
to measuring the wavelength-dependent PSF of the target. Conversely, one can
use a priori knowledge of the stellar PSF to fit this profile in any datacube
slice at the predicted centroid position, and derive a flux estimate even
in the presence of a complex and spatially variable background surface 
brightness distribution.

Considering the observational problems outlined in \S~\ref{OBSPROBL}, and
comparing with spectrographs posessing slits of any geometry, we note
that there is a number of advantages, which become important especially for 
background-limited spectroscopy of faint objects :

(1) 3D spectroscopy is insensitive to pointing and guiding errors when observing
point sources. XPN at the distance of M31 have equivalent V magnitudes fainter 
than 21. Near the nucleus, they are normally invisible for broad-band 
TV acquisition cameras on 4m-class telescopes. With 3D, the slit losses 
outlined in \S~\ref{SLITLOSSES} are no longer of concern. 

(2) Since there is no fixed physical aperture, an optimal digital aperture 
can be chosen throughout the datacube, thus optimizing the S/N at any given 
wavelength and solving the problem of \S~\ref{SEEING}.

(3) In a datacube, differential atmospheric refraction is observed as
a translation in (x,y) as a function of $\lambda$ (Arribas et al.\ 1998a).
Provided that the point source is not located too close to the edge of
the field in the direction of translation, it will always be contained in
the FOV and available for analysis, solving problem \S~\ref{REFRACTION}.

(4) A most serious drawback of slit spectroscopy is the fact that there 
are only two regions along the slit next to the source, which can be used to
estimate the background at the location of the target, setting 
e.g.\ the detection limit for Ly$_\alpha$ emitters at 8-10m class
telescopes to a few 10$^{-17}$~erg/cm$^2$/sec (Morris et al. 2002). 
This limit is dictated by systematic errors of slit geometry and 
other instrumental effects, even when the spatial distribution of the night 
sky background can be assumed to be constant (Glazebrook~\& Bland-Hawthorne 2001). 
The situation for the observation of our XPN is considerably worse: the surface
brightness of unresolved stars near the nucleus of M31 is up to 7~mag brighter
than the continuum of the night sky background, and spatially variable on
different scales. Because of the composition of the underlying stellar
population, the spatially variable surface brightness distribution may exhibit
a variance in absorption line profiles depending on the contribution from
stars of different spectral types. In addition, there is a spatially 
uncorrelated pattern of emission line spectra emerging from H$\;$II regions, 
supernova remnants, the diffuse ISM, and other nebulosities
(see Fig.~\ref{M31Halpha}).

Contrary to slit spectroscopy, 3D provides the full 2-di\-men\-sio\-nal spatial 
information at any wavelength in the vicinity of the target. We will show
that fitting a PSF to the XPN image at any wavelength slice of the 
datacube allows us to provide the best possible discrimination of the point
source against the variable background, thus reducing systematic errors which 
are  otherwise unavoidable for slit of fiber spectrographs. The details of this 
procedure are explained in \S~\ref{RESULTS}.

\section{Instrumentation}
\label{INSTRUMENT}

3D instruments have been developed on the basis of different principles: 
fiber bundles (e.g. Barden~\& Wade 1988), micro-pupil lens arrays (e.g.\
Bacon et al.\ 1995), micro-pupil lens arrays coupled to fibers (e.g.\
Allington-Smith et al.\ 1998b), and slicers (e.g. Weitzel et al.\ 1996).
PMAS is based on the fiber-coupled lens array design (see details below).
In order to prepare for the use of PMAS already during its construction 
phase, and for reasons of data reduction software development,
a preliminary observing  programme was conducted with two other existing 
3D instruments.

\subsection{MPFS}
MPFS\footnote{\url{http://www.sao.ru/\~{}gafan/devices/mpfs/mpfs\_main.htm}}
is a 3D instrument featuring the lens array~--~fiber bundle
design. It has been in operation since 1997  in the prime focus of the 6m BTA 
in Selentchuk, Russia (Si'l\-chen\-ko \& Afanasiev 2000). It has a 16$\times$16
element lens array of square lenses, with a spatial sampling of 0.5~arcsec, 
0.75~arcsec, and 1~arcsec at different magnifications, respectively.
The 256 spectra are imaged onto a thinned 1K$\times$1K SITe TK1024 CCD
using a f/1.2 Schmidt-Cassegrain camera. The total efficiency (excluding the
atmosphere) is about 5\% at 500~nm, dropping steeply towards the blue. From a 
variety of gratings, dispersions of 1.35$\ldots$5~{\AA}/pixel can be selected.

\subsection{INTEGRAL}
INTEGRAL\footnote{\url{http://www.ing.iac.es/\~{}bgarcia/integral/html/integral\_home.html}}
is a fiber bundle type of 3D spectrograph at the 4.2m WHT, La Palma
(Arribas et al.\ 1998b).
One of three different bundles with 0.45~arcsec, 0.9~arcsec, and 2.7~arcsec
projected fiber diameters and with 205, 219, and 135 fibers, respectively, 
are available to accomodate different observational requirements in terms of 
spatial resolution and FOV. The fibers are packed in a hexagonal arrangement, 
filling a rectangular field.
The fiber bundles are coupled to WYFFOS, the bench-mounted fiber 
spectrograph on the Nasmyth platform of the WHT. The detector is a thinned 
1K$\times$1K CCD at the internal focus of the spectrograph Schmidt camera.
Different gratings with dispersions between 0.35 and 6.2~{\AA}/pixel can be 
used.

\subsection{PMAS}
PMAS\footnote{\url{http://www.aip.de/groups/opti/pmas/OptI\_pmas.html}}
is a dedicated 3D instrument with a lens array of 16$\times$16 square elements 
in the present configuration, coupled by means of a fiber bundle to a fully 
refractive fiber spectrograph, which is currently equipped with a 2K$\times$4K 
thinned CCD (SITe ST002A), providing 2048 spectral bins. 
A 2$\times$2K$\times$4K mosaic CCD, which was commissioned recently, 
increases the free spectral range to 4096 spectral bins. Along the spatial
direction on the CCD, the current setup provides more than twice the space which
would be required for the total of 256 fiber spectra. The present fiber bundle
has been conservatively manufactured with 100$\mu$m diameter, high OH$^-$ doped
fibers for good UV transmission. A future upgrade with 50-60$\mu$m diameter fibers
will replace the existing IFU with a 32$\times$32 element array. The optical system
is based on UV-transparent media and provides good transmission in the blue.
A unique feature of PMAS is the internal A\&G camera, equipped with a LN$_2$-cooled,
blue-sensitive SITe TK1024 CCD, giving images with a scale of 0.2~arcsec/pixel 
and a FOV of 3.4$\times$3.4~arcmin$^2$. The camera can be used with broad-band
and narrow-band filters of the standard Calar Alto filter inventory.
For a more detailed description, see Roth et al.\ 2000a and Kelz et al.\
2003.

\begin{figure} [h]
\plotone{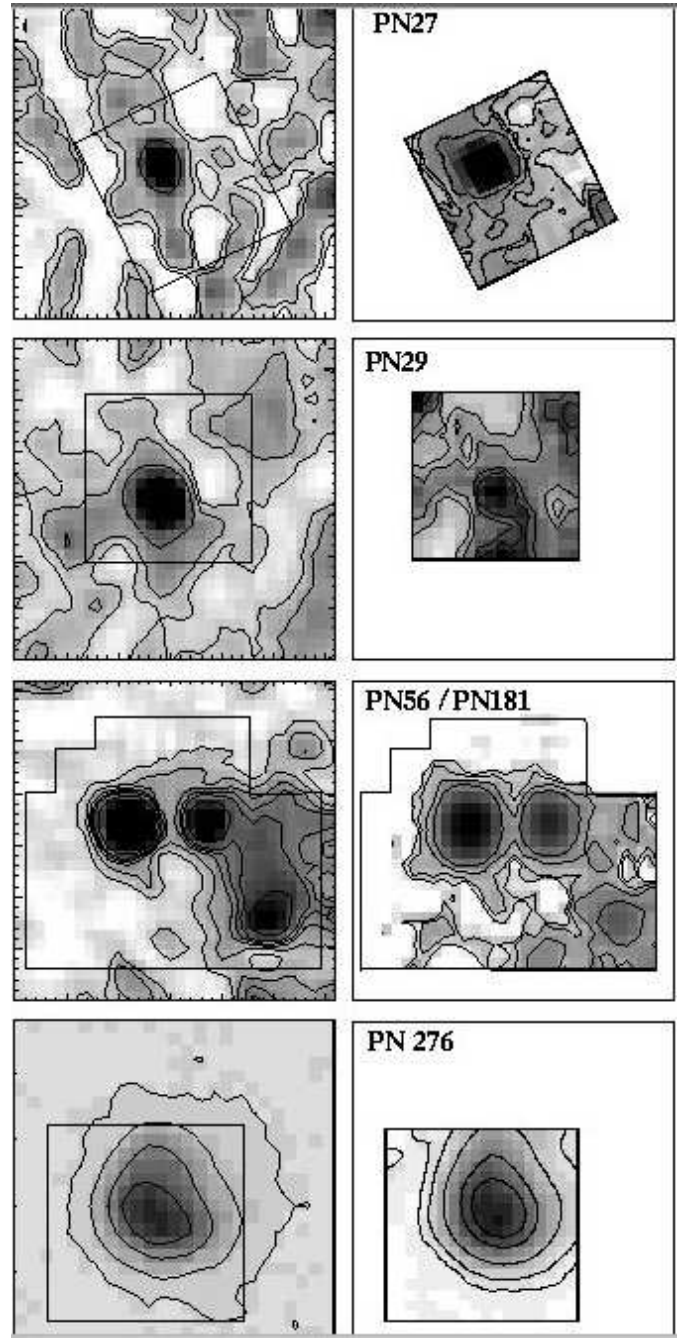} 
\caption{Calar Alto CCD images of XPN in H$_\alpha$, compared with 
reconstructed narrow-band maps from 3D data cubes. 
Left: 3.5m prime focus Fabry-P\'erot images, except for PN276 
(2.2m, CAFOS + narrowband filter).
Right: data cube maps, obtained with MPFS (PN27) and PMAS (PN29, PN56/181, PN276).
Central wavelengths are at H$_\alpha$, except for PN56/181 ([O$\;$III] $\lambda$~5007),
which had been observed only in the blue.
Note that the extended wings of the 9~{\AA} FWHM Fabry-P\'erot transmission profile 
introduce a non-negligible contamination from [N$\;$II] $\lambda\lambda$6548,6583,
which was mimicked in our 3D maps. The scale is 0.5~arcsec/pixel. 
North is up, East to the left.
\label{M31XPN_Halpha}}
\end{figure}

\section{Observations}
\label{OBSERV}
In order to understand the problems of real 3D data and for the purpose of
developing the P3d data reduction software (Becker 2002) well before the 
commissioning of PMAS, we conducted a preliminary programme of observations
with other existing instruments (MPFS, INTEGRAL). We selected our targets
from the list of CJFN89, supplemented with narrow-band H$_\alpha$
images for the purpose of finding objects which appeared to be critical in 
terms of background subtraction. The H$_\alpha$ frames were kindly made 
available to us prior to publication by T.~Soffner, University of Munich. 
The images were taken on September 10, 1994, with the Calar Alto 3.5m prime 
focus CCD camera, equipped with a Fabry-Per\'ot etalon.
The etalon has a passband of $\approx$9~{\AA} FWHM and  can be 
scanned in wavelength, thus providing flux and radial velocity information 
at the same time (Meisenheimer \& Hippelein 1992). 
An example of 3 coadded frames (exposure time 3$\times$1000sec) for a 6~arcmin 
diameter FOV,  centered on the nucleus, is shown in Fig.~\ref{M31Halpha}. 
From the on-band image, a corresponding off-band image has been subtracted, 
revealing the emission line component with the XPN as point sources, residuals 
of less than perfectly subtracted stars, and the prominent spiral-like 
appearance of the ISM emission near the nucleus and throughout the bulge of M31. 
This feature has been known for some time (M\"unch 1960). The first optical
images were obtained by Jacoby et al.\ (1985). Spectra by Ciardullo et al.~(1988)
reveal emission in H$_\alpha$ and the strong low excitation lines of 
[N$\;$II] $\lambda$$\lambda$6548,6583 and
[S$\;$II] $\lambda$$\lambda$6717,6731, which are responsible for some of the 
background subtraction problems reported by RC99. 
This fact is illustrated further with the thumbnail images of our targets in
Fig.~\ref{M31XPN_Halpha}, where the fuzzy and filamentary structure of the 
emission line background becomes quite obvious. 

A log of our observations quoting the different instruments used is given in 
Table~\ref{TBL-1}. Note that 3D spectra were taken with different wavelength 
coverage, spectral resolution, and spatial sampling, depending on the instrument 
and prevailing observing conditions. The object ID numbers were adopted from
CJFN89. Each target observation was accompanied by standard
star observations and bias, continuum and spectral line lamp calibration 
frames, which, for best accuracy, where normally taken before and after a
series of target exposures at any pointing for MPFS and PMAS, in order to
minimize flexure effects. INTEGRAL, employing the bench mounted WYFFOS 
spectrograph, is insensitive to flexure.

 \begin{table*}
 \begin{center}
 \caption{Log of Observations.\label{TBL-1}}
 \begin{tabular}{cclrccrcrr}
 \tableline\tableline
  Date        & Instrument & Object & Exposure      & Seeing   & Grating & $\Delta\lambda$ [{\AA}] & {\AA}/pixel & Flux Standard(s)     \\

 \tableline
 06-Nov-1997 & MPFS       & PN29   & 3$\times$1200 & 1.5      & V600    & 4000-6750               & 2.65       & BD+284211, Feige~24  \\
 06-Nov-1997 & MPFS       & PN276  & 2$\times$900  & 1.5      & V600    & 4000-6750               & 2.65       & Hiltner-600          \\

 18-Sep-1998 & MPFS       & PN276  & 2$\times$1200 & 1.1      & V600    & 4270-6910               & 2.6        & BD+284211, Feige~24  \\
 21-Sep-1998 & MPFS       & PN29   & 8$\times$1200 & 1.5-2.0  & V600    & 4210-6850               & 2.6        & Feige~24             \\
 27-Sep-1998 & MPFS       & PN27   & 4$\times$1800 & 2.0-2.5  & V600    & 4210-6850               & 2.6        & BD+284211            \\
 28-Sep-1998 & MPFS       & PN27   & 4$\times$1800 & 2.5-3.0  & V600    & 4210-6850               & 2.6        & BD+284211            \\

 26-Dec-1998 & INTEGRAL   & PN276  & 3$\times$1800 & 1.5      & R600B   & 3540-6660               & 3.01       & SP0305+261, G191-B2B \\

 23-Oct-2001 & PMAS       & PN276  & 3$\times$1200 & 1.5      & V600    & 3550-5200               & 1.6        & HR718                \\
 24-Oct-2001 & PMAS       & PN29   & 9$\times$1200 & 1.1      & V600    & 5600-7200               & 1.6        & HR153, HR1544        \\
 26-Oct-2001 & PMAS       & PN29   & 8$\times$1800 & 1.4      & V600    & 3550-5200               & 1.6        & HR153, HR1544        \\
 31-Aug-2002 & PMAS       & PN56   & 4$\times$1800 & 1.0      & U600    & 3440-5100               & 1.6        & HD192281             \\
 01-Sep-2002 & PMAS       & PN56   & 9$\times$1200 & 1.1      & U600    & 3440-5100               & 1.6        & HD192281, HD19445    \\
 01-Sep-2002 & PMAS       & PN181  & 5$\times$1200 & 1.1      & U600    & 3440-5100               & 1.6        & HD192281, HD19445    \\

 \tableline
 \end{tabular}
 \end{center}
 \end{table*}

\section{Data Reduction}
\label{DATARED}

The data reduction for all observations, except for MPFS data of 1997, was 
performed using the P3d package of Becker (2002). P3d is an IDL code which
was originally developed for PMAS, but has been adapted to 3D data from other 
instruments (SPIRAL, INTEGRAL). The 1997 data were reduced by JS using
a modified version of the {\em dofiber} task of IRAF  \footnote{IRAF is distributed 
by the National Optical Astronomy Observatories, which is operated by the 
Association of Universities for Research in Astronomy, Inc. (AURA) under 
cooperative agreement with the National Science Foundation.}. This task
involved a considerable amount of interactive work and required to control
each of the $\approx$256~spectra individually for any single CCD frame, whereas
the P3d programme works automatically once a set of parameters has been
set appropriately. The various steps of the data reduction procedure are as
follows. The raw CCD frames are first bias subtracted and cleaned from cosmic 
ray events. As an option, P3d allows to apply a correction for pixel-to-pixel
CCD response non-uniformity, known as the standard flatfield correction for direct 
imaging applications. This correction is performed using grossly defocussed
continuum flatfield exposures and corrects only for small scale pixel-to-pixel
variations. Note that for 3D fiber spectroscopy, there are different
levels of response calibration: (1) detector pixel non-uniformity, (2) {\em total} 
spaxel-to-spaxel throughput variation (caused by lens array defects and 
diffraction, variation in fiber transmission and coupling), and (3) the 
{\em wavelength-dependent} non-uniformity of spaxel-to-spaxel response, which is 
also linked to the wavelength-dependent variation of the detector QE. 

After these first basic procedures, a critical step consisted in the 
tracing of spectra which became difficult when there was little flux, and thus
confusion for automatic procedures, in particular near the edge of a spectral
range when the response was dropping steeply. This is where the {\em dofiber} routine
normally required to correct interactively for misidentifications and errors.
P3d includes a geometrical model based on calibration exposures which is 
cross-correlated and matched with the observed set of spectra for any single
exposure in order to avoid this complication. For the case of PMAS, Fig.~4 
in Roth et al.\ 2002b shows how these spectra are arranged on the detector, 
with comfortably wide inter-order gaps (14 pixels distance), which is an
unusual arrangement compared to most other 3D instruments. In order to make 
efficient use of detector estate, one would normally try to squeeze  as many 
spectra as possible onto the CCD. The inevitable overlap of adjacent
spectra to a degree which depends on the distance of spectra and optical
quality of the spectrograph optics leads to ``crosstalk'' with effects as
discussed by Roth et al.\ 2000b, Becker et al.\ 2000, and investigated 
in detail by Becker (2002); but see also Allington-Smith~\& Content 1998. 
We shall further discuss this issue below.

P3d allows for two different modes of extraction, which is the next step in
the data reduction process. The standard mode is to define an extraction 
swath along the traces and integrate the contained flux for any spectral bin. 
This is the procedure which is also employed in a subset of P3d routines, 
forming the PMAS online data reduction software for use at the telescope. 
Swath extraction fails badly when the spacing of spectra is very dense, in 
particular for applications where one is interested in faint spectral signatures 
on a bright background. Such features are likely to be completely wiped out 
in the presence of crosstalk. This difficult situation was encountered with 
our MPFS data sets.
In order to correct for crosstalk, P3d provides a profile-fitting mode
which can be applied when a model for the cross-dispersion profile can
be constructed. The details of this procedure are described in a separate
paper (Becker, in preparation). Profile-fitting also allows to correct
for scattered light, which has been a considerable complication for our MPFS 
spectra in the blue (Fig.~4 in Roth et al.\ 2000b).

After extraction, a wavelength calibration is applied using spectral line lamp 
exposures. This is another step where MPFS with {\em dofiber} required human 
interaction and correction for errors. P3d uses calibration lamp template
spectra resulting in a robust procedure with no need for interactive
corrections.

The set of extracted and wavelength calibrated spectra was then corrected
for spaxel-to-spaxel (or fiber-to-fiber) variations by normalizing to a
continuum flatfield exposure. A final correction of the smooth wavelength-dependent
spaxel-to-spaxel response variation of order 3-5\% rms over the entire
wavelength range was provided by means of a twilight sky flatfield exposure 
for the MPFS (1998) and PMAS frames (``fiber flat''), omitting any disturbing
atmospheric features. Using standard star exposures, the set of spectra was
then  weighted with the instrumental and atmospheric sensitivity function to 
produce the final flux-calibrated datacube.

The resulting dataset consists of a 2-dimensional frame of stacked spectra
which we found to be a convenient arrangement for inspection with a visualisation
tool. An example for the appearance of this tool, which allows one to plot maps, 
single or coadded spectra for mouse-selected digital apertures, is shown in Roth
et al.\ 2002c. The data was finally written to disk in a FITS compatible format
whose specification was developed by the Euro3D consortium
(Walsh \& Roth 2002).

\section{Data Analysis and Results}
\label{RESULTS}

This section is organized as follows: in \S\S~\ref{PSF} -- \ref{PSF3DSPEC} 
we discuss general properties of our reduced 3D data, the procedure of merging 
single exposures into a final datacube, our techniques of background subtraction,
3D spectrophotometry, and the final steps to derive the flux-calibrated and 
dereddened XPN spectra. In \S\S~\ref{PN29} -- \ref{PN276} we present the results 
for individual objects. If not indicated otherwise, the software used for the 
various steps was developed under IDL and written by TB.

\subsection{3D Point Spread Function}
\label{PSF}

\begin{figure}[h]
\plotone{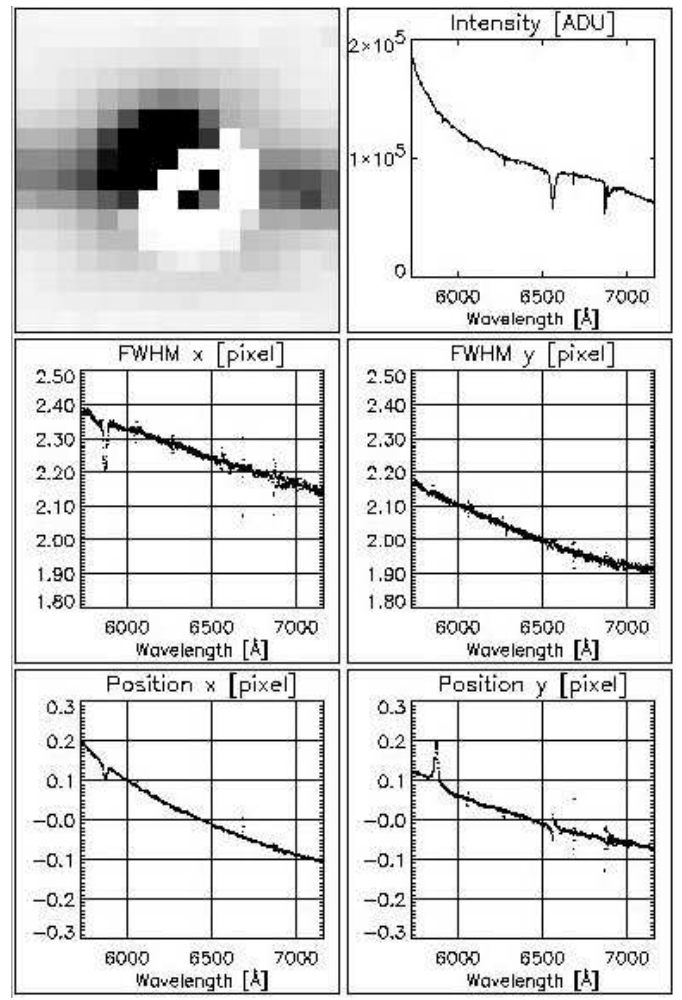}
\caption{Moffat function PSF analysis of standard star datacube slices. The upper 
left panel shows the mean residual of the fit, with a greyscale stretch
from 0\% (white) to +1\% (black) of peak intensity. For explanation see text.
\label{PSFANALYSIS}}
\end{figure}

A crucial test as to whether or not the proposed PSF fitting technique would be
useful for 3D data consisted in the analysis of standard star datacubes, which
were well-enough exposed to provide good S/N in any quasi-monochromatic slice
(Fig.~\ref{PSFANALYSIS}, upper right plot).
The slices were treated like normal, albeit tiny, CCD images, and a Moffat function
was fitted independently to the frame of each wavelength bin. As an instructive
example, the result for a particularily {\em poor} exposure of the standard star 
HR1544
is shown in Fig.~\ref{PSFANALYSIS}, chosen for the purpose of demonstrating that 
it is indeed possible to determine an accurate and well-defined PSF. The
datacube was obtained with the P3d\_online data reduction routine without
profile fitting. In the two panels in the middle we plot the 
resulting FWHM fit values in X and Y vs.\ wavelength, and in the bottom panels the
corresponding centroid offsets in X and Y with respect to a wavelength in the 
middle of the free spectral range. A map of the mean residual of the PSF-fit is 
displayed in the upper left panel. 

First of all, we note that the centroid is not located
at a fixed position, but moves monotonically with wavelength in X and Y. The
standard star was observed at an hour angle of 1$^h$~04$^m$ and at airmass 1.177,
with an expected offset due to differential atmospheric refraction of 0.18~arcsec
between the wavelengths of 5700~{\AA} and 7300~{\AA}, according to Filippenko
(1982). The ambient parameters used for this estimate are a temperature of 
9.5$^\circ$C, a relative humidity of 33\%, and an atmospheric pressure of 789.8 mbar.
The same offset of 0.18 arcsec is obtained in our datacube when we add the X and Y
offsets in quadrature (a spatial element corresponding to 0.5~arcsec). 

Secondly, it is seen that the scatter about a mean curve is very low: a best fit 
low order polynomial yields an rms scatter of 0.0015 and 0.0018 spaxels in X and Y,
respectively, indicating that the PSF is an extremely sensitive measure of position 
($\pm$0.001 arcsec). The obvious flaw near 5800~{\AA}, introducing a systematic 
error of $\le$0.1~arcsec  over a  small wavelength interval, is entirely due to 
a local detector blemish which could not be flatfielded out. The flaw vanishes 
completely in exposures where the star is centered on another region of the IFU. 
Likewise, we show that similar systematic errors at a few isolated wavelength bins 
near the cores of the H$_\alpha$ and other absorption lines are caused by small 
differential wavelength calibration errors between different sets of spectra
and the effects of spectral rebinning.

Thirdly, we observe that the PSF is non-symmetric as seen in the residual image 
and FWHM plots, indicating an elliptical elongation, and an extended horizontal 
feature at a level of 1~\% peak intensity. This result is instructive in two ways: 
(a)~the horizontal feature illustrates the effect of crosstalk, introduced because 
of the presence of the extended wings of the fiber spectrograph PSF. The effect 
amounts to typically 0.5\% peak intensity and vanishes when the profile-fitting 
extraction is applied; 
(b)~the elongation indicates an optical aberration of the 3.5m Telescope, which was 
in fact identified as decentering coma and astigmatism due to a misalignment of the 
primary mirror, and which was corrected during a maintenance period shortly after 
our August~2002 run (Thiele 2003).

Finally, we note that the expected decrease of seeing disk size with wavelength 
is clearly seen in the plots of FWHM (Fried 1966).

\subsection{Co-adding Datacubes, Atmospheric Refraction}
\label{3DMOSAIC}

The measurement of faint line intensities of XPN at the distance of M31 using
a 4m-class telescope requires total exposure times of order several hours.
The combination of datacubes taken over such a period of time is complicated
by the fact that each exposure is affected by a different amount of differential
atmospheric refraction, i.e.\ shift of monochromatic slices with wavelength. To
counteract this effect, we shifted the slices at the wavelength of interest
to a common reference system, which was compensated for refraction, using the
formulae of Filippenko (1982). Moreover, the effect of a different 
amount of atmospheric extinction as a function of
airmass had to be taken into account when coadding all datasets to form the
final datacube. The set of PMAS exposures taken in October 2001 was suffering
from flexure of the telescope guiding camera, which required yet another 
adjustment. The net effect of flexure, which was confirmed independently with two 
different daytime tests at the telescope, resulted in an overall  shift of 
order 0.5-1~arcsec per hour, thus requiring to shift these datacubes to 
compensate both for differential refraction and flexure. Note that this 
problem has disappeared after commissioning of the PMAS A\&G camera, which is 
internal to the instrument and replaces the telecope TV-Guider system.
The necessity to correct for the shifts, however, was an interesting exercise 
for 3D mosaic exposures of areas larger than provided by the FOV of the IFU.
The maps shown in Fig.~\ref{M31XPN_Halpha} therefore cover a larger area than 
normally obtained from the instrumental FOV alone.

We note in passing, that the continuous offsets between the exposures resemble 
the familiar {\em dithering} technique for direct imaging, with the potential
advantage of improving the sampling of the PSF, and reducing residual errors
of the spaxel-to-spaxel non-uniformity correction (Wisotzki et al.\ 2003).

\subsection{3D Spectrophotometry}
\label{PSF3DSPEC}

The last step to create the spectra of our targets consisted in measuring
the flux within a digital aperture or using a PSF fit around the centroid of 
the fully corrected, final datacube at all wavelengths. 

In order to do this accurately enough, one has to subtract a two-dimensional 
model of the background surface brightness distribution, since contrary to 
the night sky emission, this light is anything but constant over the field-of-view.
In our case, the background consists of the two dominant components: stellar
continuum, and diffuse/filamentary ISM emission. As discussed in more detail 
in \S~\ref{DISCUSSION}, the latter component presented a peculiar problem, since
it is difficult to distinguish from the continuum, being responsible for intensity 
overestimates of some weak XPN emission lines in conventional spectrophotometry.

A two-step procedure was devised to accomplish the modelling and subsequent removal
of these components. The XPN case (as opposed to stars) is quite favourable, in
that the object continuum (typically 10$^{-3}\times$I$_{H_\beta}$/{\AA}) is completely 
negligible in comparison with the background intensity of the galaxy .
The entire datacube consists essentially of background light, except for the 
few wavelengths with appreciable XPN emission line flux. Using a normalization scheme
for all of the monochromatic slices to take out the continuum spectral features, it was
possible to (1) create continuum maps as a function of wavelength, (2) enhance the 
contrast of the ISM emission line filaments, and (3) create a map of the emission 
component. The background correction was finally completed by subtracting the two
components from the original datacube. The technical details of this procedure
are described in Becker (2002).

Although we cannot completely rule out the possibility that at the wavelengths of 
strong background absorption  or emission line features the chance alignment of 
PSF-compatible anomalies with the XPN, i.e.\ unrelated point sources, 
would remain undetected and cause systematic 
errors of the XPN line fluxes,  we expect that the probability of such chance 
alignments is small. In general we find that our continuum subtraction yields 
flat spectra which are free of  obvious residuals (Fig.~\ref{XPNSPECTRA}), 
contrary to the examples discussed in  \S~\ref{PREVIOUS}. 

Because of the relatively high contrast of the [O$\;$III] $\lambda$5007 line 
in an almost featureless region of the background continuum, 
the centroid was defined to an accuracy of a fraction of a spaxel 
(e.g.\ 0.1~arcsec in the case of PN29), and the flux was obtained directly 
from aperture photometry. Using a standard star curve-of-growth analysis, 
an aperture correction was applied to account for the flux lost outside of 
the digital aperture. The PSF at this wavelength
was determined as a 2-dimensional gaussian and used as a template for the 
other, fainter emission lines with a correction for the wavelength-dependent 
variation of the seeing FWHM. Since the measurements are background shot-noise 
dominated, a more sophisticated PSF characterization was unnecessary.

The final procedure was then to fit the wavelength-dependent
model PSF to each datacube slice, constrained by the centroid of the bright
[O$\;$III] feature. Note that the fit allows for negative fluxes, which, in the
presence of noise, avoids a positive bias of the detection probability. Note
also that the absence of any limiting apertures (slit effects) justifies the
term {\em spectro\-photo\-metry}.

\subsection{2-Channel Deconvolution}
\label{CPLUCY}
As a more sophisticated method to separate a point source from the background, we used
the two-channel Richardson-Lucy deconvolution algorithm {\em cplucy} (Lucy 1994) which 
is available within IRAF. It is optimized for improved photometric performance and 
avoids the artifacts and oscillations that are known for the standard 
RL-Algorithm (Hook \& Lucy 1993). The two-channel method makes use of the
precise knowledge of the location of one or several point sources, and iterates
the point source channel separately from another channel, which is processing 
the smoothly varying background. We applied the algorithm to our datacubes such that
each monochromatic slice was subject to an independent {\em cplucy} run. The PSF is defined 
by one of the bright lines, e.g.\ [O$\;$III] $\lambda$5007, corrected for the
weakly varying FWHM as a function of wavelength as determined from standard star 
exposures. The flux of the point source detection in each slice in the end created 
the final spectrum. Details  of the application of the two-channel deconvolution 
method to the peculiar case of a datacube, the treatment of a photometric bias in 
the case of very low background intensity levels, and various tests are described 
in more detail in Becker (2002).

As a drawback, we note that the procedure of fitting an optimized 
PSF to the centroid of an XPN in each wavelength bin (i.e.\ monochromatic map), is 
forced to produce detections with flux~$\ge 0$. The resulting spectra have non-normal 
noise distributions. They were preferrentially used for confusion-limited cases and to 
determine upper limits for faint, background-limited lines.

\subsection{Dereddened Line Intensities}
\label{DEREDDEN}

The emission line intensities were measured after flux-calibra\-ting the
extracted XPN spectra with standard star exposures by fitting gaussians at 
the nominal wavelengths, corrected for the radial velocity shift known from 
the bright [O$\;$III] line. This step was performed within the ESO-MIDAS 
data reduction package and through use of a line fitting tool kindly provided
by A.~Schwope (AIP). We have assessed the statistical errors of these fits
by implanting artificial emission lines with known intensities at several 
(typically $\approx$100) different wavelengths near the original line into 
the data, fitting these simulated lines, and evaluating the standard error 
of a single measurement from the variance of the whole set.
The dereddening based on the Balmer decrement followed the standard procedure
as described by RC99.

The final results for our objects are listed in Tab.~\ref{TBL-2}. The errors
for typical line intensities (faint/intermediate/bright lines) are listed
as S/N values at the end of the table.


 \begin{figure*}[t]
 \centering\includegraphics[width=1.0\linewidth,clip]{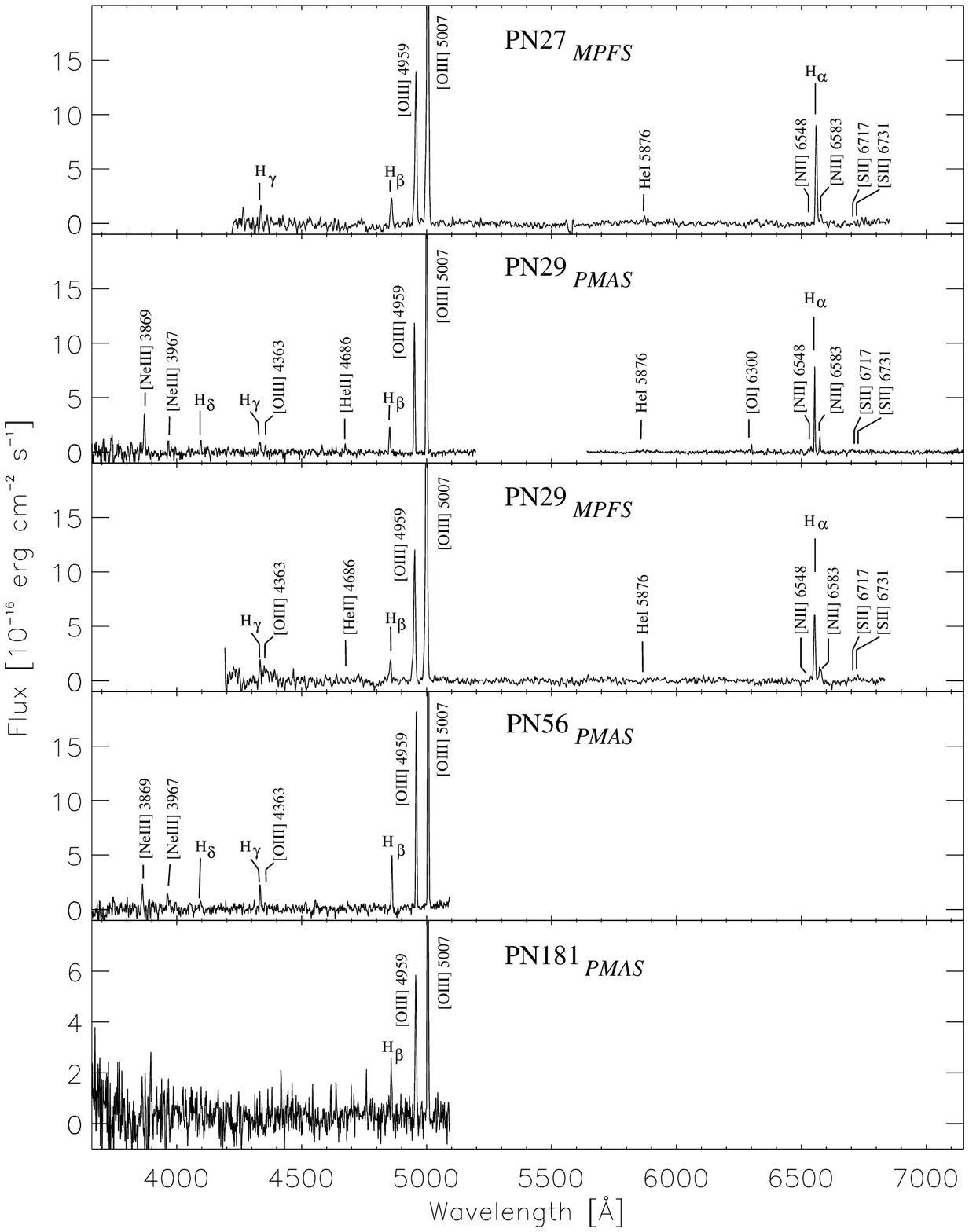}
 \figcaption{Fully reduced and flux-calibrated spectra of our XPN sample.
 The spectral bin width for each spectrum [{\AA}/pixel] is listed in Tab.~\ref{TBL-1}.
 \label{XPNSPECTRA}}
 \end{figure*}


 \begin{figure*}[htb]
 \centering\includegraphics[width=1.0\linewidth,clip]{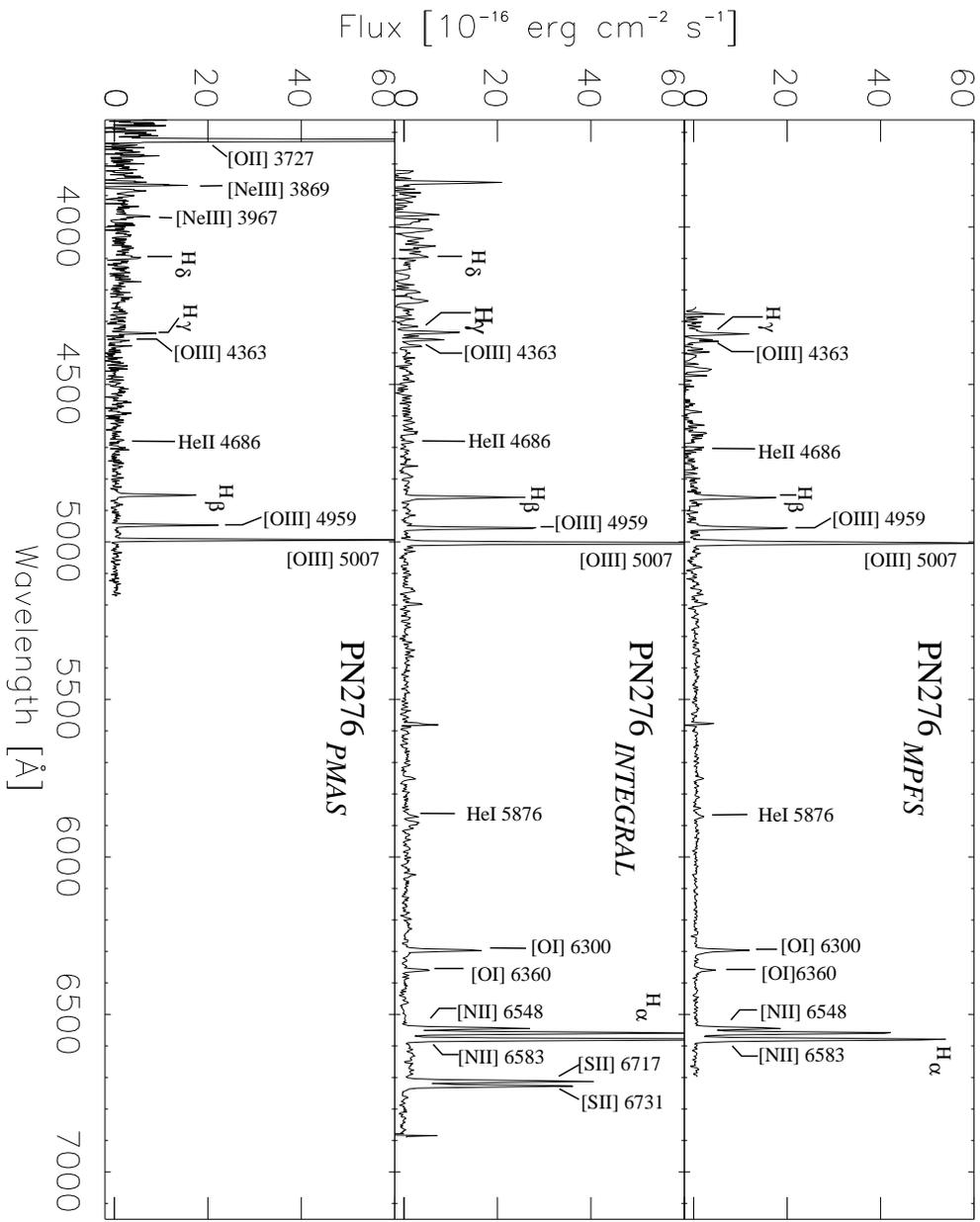}
 \figcaption{Spectra of the SNR PN276, observed with MPFS, PMAS, and
 INTEGRAL. The spectral bin width for each spectrum [{\AA}/pixel] is listed in 
 Tab.~\ref{TBL-1}.
 \label{SNRSPECTRA}}
 \end{figure*}

 \begin{table*}[h]
 \begin{center}
 \caption{REDDENING-CORRECTED LINE STRENGTHS.\label{TBL-2}}
 \begin{tabular}{clrrrrrrrr}
 \tableline\tableline

 $\lambda$ [\AA] &  Ion  &  PN27 & PN29 & PN29 & PN56 & PN181 & PN276  &  PN276  &  PN276 \\
            &       &  MPFS & MPFS & PMAS & PMAS & PMAS  & MPFS   &  INT.   &  PMAS  \\
 \tableline
 3727  &  [O$\,$II] & ...   & ...  & <26  & <38  & <83   & ...    &  ...    &  820   \\
 3869  & [Ne$\,$III] & ...  & ...  & 145  & 26   & 95    & ...    &  84     &  71    \\
 3889  & He~I, H~I  & ...   & ...  & <15  & <29  & <74   & ...    &  ...    &  29    \\
 3969  & [Ne$\,$III], H~I &...& ...& 56   & <26  & <71   & ...    &  32     &  40    \\
 4101  & H$_\delta$ & ...   & ...  & 40   & 17   & <67   & ...    &  ...    &  31    \\
 4340  & H$_\gamma$ & 47    & 48   & 43   & 46   & <64   & 49     &  51     &  47    \\
 4363  & [O$\,$III] & 37    & ...  & 19   & 27   & <64   & 19     &  29     &  12    \\
 4686  &  He~II     & ...   & 23   & 19   & <19  & <62   & ...    &  ...    &  11    \\
 4861  & H$_\beta$  & 100   & 100  & 100  & 100  & 100   & 100    &  100    &  100   \\
 4959  & [O$\,$III] & 542   & 612  & 565  & 399  & 368   & 117    &  122    &  118   \\
 5007  & [O$\,$III] & 1645  &2030  & 1698 & 1282 & 1121  & 365    &  376    &  345   \\
 5755  &  [N$\,$II] & ...   & ...  & <5   & ...  & ...   & 10     &  10     &  ...   \\
 5876  &  He~I      & 17    & 12   & 12   & ...  & ...   & 11     &  12     &  ...   \\
 6300  &  [O$\,$I]  & ...   & ...  & 10   & ...  & ...   & 69     &  83     &  ...   \\
 6360  &  [O$\,$I]  & ...   & ...  & <5   & ...  & ...   & 24     &  26     &  ...   \\
 6548  &  [N$\,$II] & ...   & 16   & 15   & ...  & ...   & 124    &  157    &  ...   \\
 6563  & H$_\alpha$ & 286   & 286  & 286  & ...  & ...   & 304    &  310    &  ...   \\
 6583  &  [N$\,$II] & 13    & 48   & 35   & ...  & ...   & 380    &  381    &  ...   \\
 6678  &  He~I      & ...   & ...  & 15   & ...  & ...   & ...    &  ...    &  ...   \\
 6717  &  [S$\,$II] & ...   & 18   & <5   & ...  & ...   & 193    &  180    &  ...   \\
 6731  &  [S$\,$II] & ...   & 30   & <5   & ...  & ...   & 188    &  176    &  ...   \\
 7065  &  He~I      & ...   & ...  & <5   & ...  & ...   & ...    &  ...    &  ...   \\
 7135  & [Ar$\,$III]& ...   & ...  & 12   & ...  & ...   & ...    &  ...    &  ...   \\
 \tableline\tableline
       & Uncertainties &(in \%): \\
 \tableline
 4363  & [O$\,$III] & ...   &  ... & 14.3 &  9.4 & ...   & ...    &  ...    & 19.6   \\
 4861  & H$_\beta$  & ...   &  9.1 &  3.3 &  2.5 & 6.6   & ...    &  ...    &  2.3   \\
 5007  & [O$\,$III] & ...   &  0.7 &  0.6 &  0.4 & 0.8   & ...    &  ...    &  0.7   \\
 \tableline\tableline
 m$_{5007}$&        & 20.81 & 20.95& 21.01 & 20.96:& 21.88:& 19.99:& 20.13: & 20.45  \\
 c     &            & 0.34  & 0.41 &  0.30 & 0.33  & 0.33  &  ...  &  ...   &  ...  \\               
 \tableline
 \end{tabular}
 \tablecomments{Dereddened line intensities are tabulated in units of I(H$_\beta$)=100. 
   The instrument used for each measurement is indicated in row 2. Typical uncertainties
   for lines of low, intermediate, and high intensity  as derived from fitting emission 
   line profiles to [O$\,$III] $\lambda$4363, H$_\beta$, and [O$\,$III] $\lambda$5007 are 
   listed in the lower part of the table. Faint lines for which we did not succeed to
   obtain a reliable fit are indicated with 5$\sigma$ upper limits.
   The last two rows give the measured monochromatic magnitudes m$_{5007}$, 
   and the logarithmic extinction c. The discrepancies in m$_{5007}$ among the
   different PN276 measurements are attributed to non-photometric conditions during the
   MPFS and INTEGRAL observations.}
 \end{center}
 \end{table*}

\section{Discussion}
\label{DISCUSSION}

\subsection{PN29}
\label{PN29}
This is the best-studied object of our sample. It was observed in 3 different
observing runs with the MPFS and PMAS instruments. We had selected PN29 initially 
for the 1997 MPFS run because of its relatively high brightness among the 
CJFN89 XPN (m$_{5007}$ = 21.01), being located nevertheless close enough to the  
nucleus of M31 to study the systematic effects of different background subtraction 
techniques (132~arcsec north-east from nucleus).
Coincidently, PN29 is one of the objects in common of the JC99 and RS99 samples,
where the two sets of measurements produced conflicting results, which can now be 
reconciled with our new 3D observations. JC99 give a detailed account of their
measurement and the difficulties to match the observed line intensities with
a physically reasonable photoionization model. The data presented two 
major problems: 
firstly, their [O$\;$III]~$\lambda$5007/H$_\beta$ line ratio of 14.8 
is significantly different from the value of 22.2 reported by RS99. 
The discrepancy is completely incompatible with the respective error estimates, 
so that the difference remained a mystery. 
Secondly, the [S$\;$II] line ratio I($\lambda$6717)/I($\lambda$6731) with a value 
of 1.76 is at the low-density limit, which required to invoke
a two-zone ionization model in order to match the observed line intensities.
Since, as a result, the sulfur abundance of this model was unusually high 
([S/H]=+0.5), the authors considered the possibility of an artifact due to the
weakness of the sulfur lines and the difficult background subtraction.


 \begin{figure}[h]
 \plotone{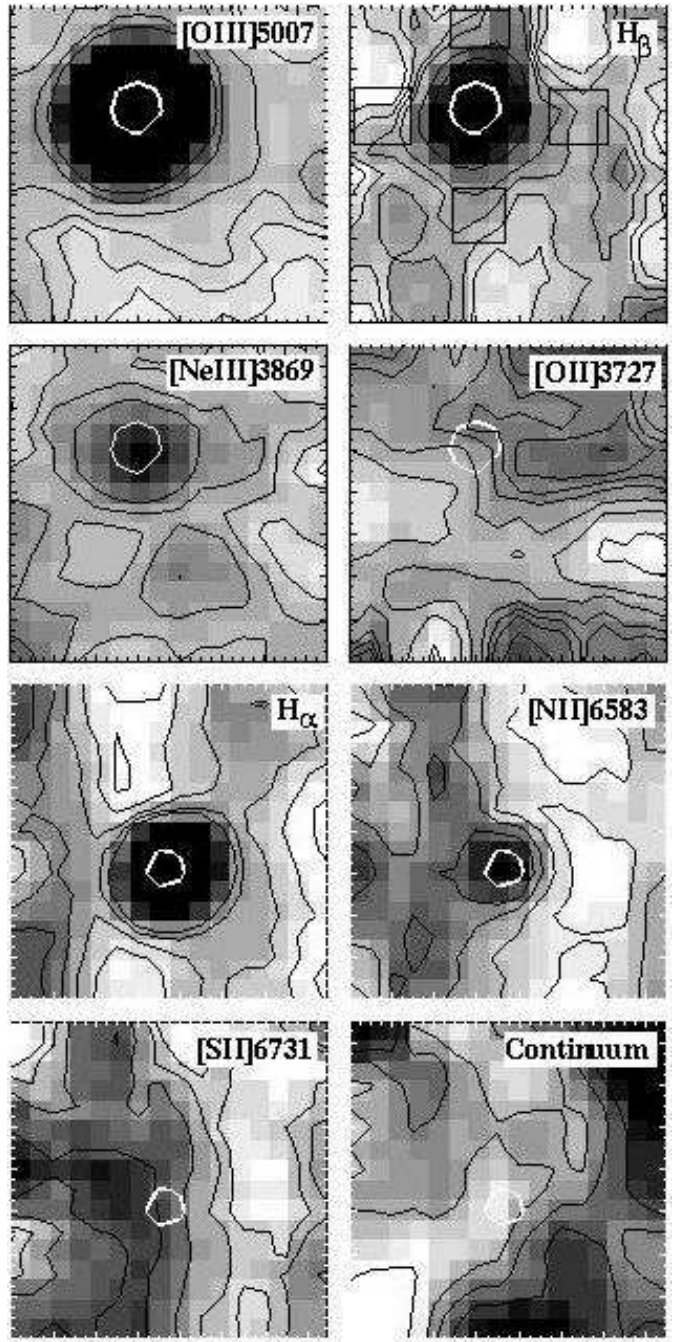}
 \caption{Monochromatic maps of PN29 in [O$\;$III], H$_\alpha$, H$_\beta$, in the 
 continuum and several low/high excitation forbidden emission lines. The positional
 offset between the upper and lower four maps is due to two different pointings, 
 taken in the blue and in the red, respectively. Except for the lower right panel, 
 the continuum has been subtracted. The overplotted centroids of the corresponding PSFs 
 in H$_\alpha$ and [O$\;$III]~$\lambda$5007 (white contour) are indicated to guide the 
 eye, in particular for the faint emission lines. The greyscale and contour levels 
 are set to arbitrary logarithmic scales in order to enhance faint features in the 
 background surface brightness distribution (white=faint,black=bright).
 Scale: 0.5~arcsec/spaxel. Orientation: N up, E left.
 \label{PN29MONO}}
 \end{figure}

\begin{figure}[h]
\epsscale{1.0}
\plotone{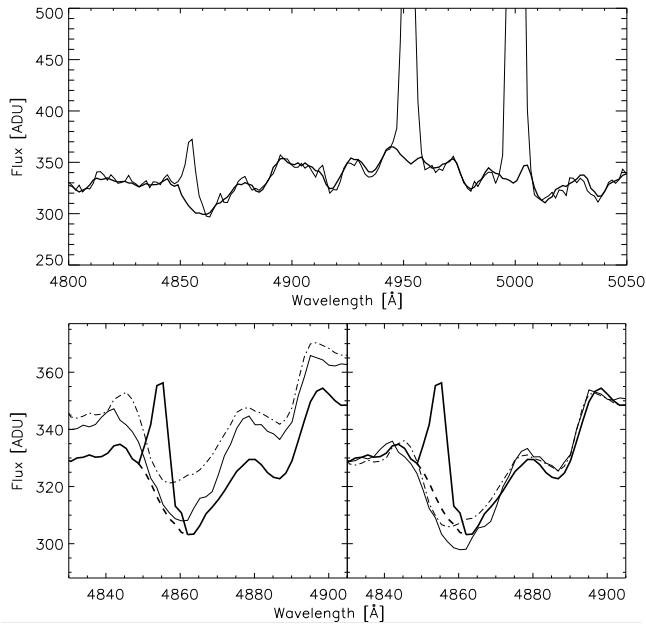}
\caption{Background subtraction in H$_\beta$. Top panel: overview of a
spectral region near 5000{\AA}. Emission line spectrum of PN29 (thin line), 
superimposed on the background continuum of unresolved stars (thick line). 
Bottom panels: simulating slit spectroscopy --- lower left: coadded flux from 
3$\times$3 spaxels centered on the XPN (thick line), background interpolation from 
simulated slit with horizontal (thin full line) and vertical orientation (
dash-dotted line).
Lower right:  the same spectra after renormalization to match the continuum 
near H$_\beta$, note the dicrepancies in the core of the absorption line. 
The dashed line is the continuum as determined from the two-channel cplucy 
algorithm.
\label{PN29HBETA}}
\end{figure}

Concerning the first problem, an example of our new data is shown in Fig.~\ref{PN29HBETA}. 
The  spectrum was obtained from the combined datacube of a total of four 1800sec exposures
using a 600~lpmm grating (1.65~\AA/pixel) by co-adding 4 spectra within a square 
digital aperture of 1$\times$1~arcsec$^2$. Next to [O$\;$III] $\lambda$5007 and 
$\lambda$4959, H$_\beta$ is seen in emission on the blue wing of the stellar 
background absorption line. The thick line is the spectrum of the background
alone, derived as the average from an annulus of 50 spaxels of radius=3~arcsec around 
the central aperture and scaled to a best match of the target spectrum. Due
to the large number of contributing spectra, the background spectrum has very little 
noise. Although at first glance the agreement with the continuum of the XPN aperture 
appears to be satisfactory, a closer inspection reveals that a measurement of the
XPN H$_\beta$ line intensity is sensitively depending on the details of the background
definition. Obviously, in order to arrive at an accurate line intensity estimate,
it is necessary to properly account for the slope of the absorption line profile
in the process of background subtraction. 

We have simulated the background subtraction of conventional slit spectroscopy by coadding
flux in virtual horizontal and vertical slitlets adjacent to the point source, as indicated 
by the rectangles in the H$_\beta$ map of Fig.~\ref{PN29MONO}. 
Linear interpolation within each slitlet pair yields the spectra shown in the lower 
left panel of Fig.~\ref{PN29HBETA} (thin full line: horizontal, dash-dotted line: vertical slit). 
By way of inspecting the continuum map (Fig.~\ref{PN29MONO}), the offset between the two 
curves is readily explained:
the vertical slitlet pair is sampling, on average, a somewhat brighter background
surface brightness than the horizontal slit does, while the XPN centroid is located near
a local minimum of the continuum background surface brightness distribution. Therefore,
the slit-based values are overestimating the background at the location of the point source, 
which is probably also why the JC99 continuum spectrum of PN29 exhibits an overall 
negative offset from zero. An attempt to bring the two former curves into agreement by 
scaling them to a best match with the source spectrum outside of the H$_\beta$ feature,
however, fails in the core of the absorption profile: 
the vertical slit predicts a significantly shallower absorption than the horizontal
slit does (lower right panel in Fig.~\ref{PN29HBETA}).
The difference can be made plausible by realizing that the background is increased 
by a diffuse emission component, which partially fills in the absorption profile 
(H$_\beta$ map in Fig,~\ref{PN29MONO}) -- a familiar
problem known also from integrated light population studies (cf.\ remarks in 
\S~\ref{INTRODUCTION}).
In our exercise, the horizontal slitlet pair is less effected by the ISM filament 
than the vertical one, thus producing the deeper absorption trough. Numerically,
the difference translates into [O$\;$III]~$\lambda$5007/H$_\beta$ line ratios of
12.0 and 19.5, respectively.
 
Although it is unlikely that our simulation is capable of reproducing {\em exactly} the 
measurements of JC99 and RS99, respectively, we can offer a sensible explanation for the 
discrepancy between their H$_\beta$ results: most likely the difference was caused
by a different setup of the slit orientation.
From this experiment it becomes clear that the simple slit approach is insufficient 
to provide an accurate  background model as required for a reliable determination of 
the contaminated XPN  H$_\beta$ line intensity. 
In contrast, the result of our PSF fitting procedure with {\em cplucy} is plotted
as the thick line in the bottom panels of Fig.~\ref{PN29HBETA}, the dashed line
indicating the background channel alone. The [O$\;$III]~$\lambda$5007/H$_\beta$ 
line ratio from this analysis is 17.0. An important consequence of the smaller
H$_\beta$ intensity compared to JC99 is an increase of our value for the logarithmic 
extinction from c~=~0.14 to c~=~0.30.

Concerning the problem of the [S$\;$II] doublet $\lambda$$\lambda$6717,6731, which 
appeared to be unusually bright (I($\lambda$6717)=51) in the data of JC99, these
lines are not even detected in our spectrum. They must be significantly fainter than
this value, with a conservative upper limit of I($\lambda$6717)$\le$5. 
A closer inspection of the [S$\;$II] map in Fig.~\ref{PN29MONO} shows that the XPN 
is embedded in a fuzzy structure of background emission, which most probably has caused 
the overestimate suspected by JC99. We note in particular that at the location
of the PSF centroid determined from H$_\alpha$, no local maximum is visible in [S$\;$II].
Furthermore, our map in [O$\;$II]$\lambda\lambda$3727,3729 also does not reveal
a trace of any local maximum, contrary to the nearby [Ne$\;$III]$\lambda$3869, which
is well-defined. 
The 5$\sigma$ upper limit for [O$\;$II] is I($\lambda\lambda$3727,3729)$\le$26.

 \begin{figure}[h]
\plotone{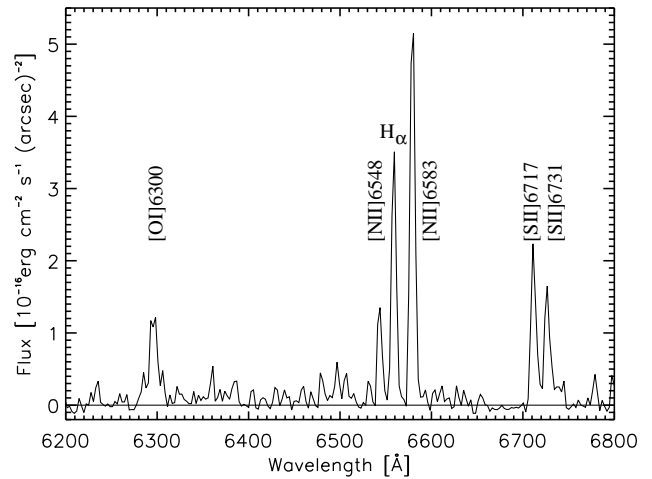}
\caption{Continuum-subtract background spectrum near PN29.
\label{PN29BACKGR}}
\end{figure}

These findings do not come entirely as a surprise.  The ISM spectra of M31 presented
by Ciardullo et al.\ 1989 show strong emission in the low excitation lines of
[S$\;$II]$\lambda$$\lambda$6717,6731 and [N$\;$II]$\lambda$$\lambda$6548,6583, with
line strengths comparable to H$_\alpha$. 
After careful modelling and subtraction of the continuum, we have produced a 
flux-calibrated spectrum of the gaseous emission background component
(Fig.~\ref{PN29BACKGR}), which compares well with the result of Ciardullo et 
al.\ 1989. Note that the order of magnitude of the [S$\;$II]$\lambda$6717
flux within a 2'' slit is compatible with the PN29 line intensity of 
4$\times10^{-16}$erg/cm$^2$/sec measured by JC99, which the authors had 
suspected to be affected by background contamination. We also confirm that the
discrepant [S$\;$II]$\lambda$$\lambda$6717,6731 line ratio of 1.76 is due to
the background filament, and not intrinsic to PN29.

\begin{figure}[t]
 \centering\includegraphics[width=1.0\linewidth,clip]{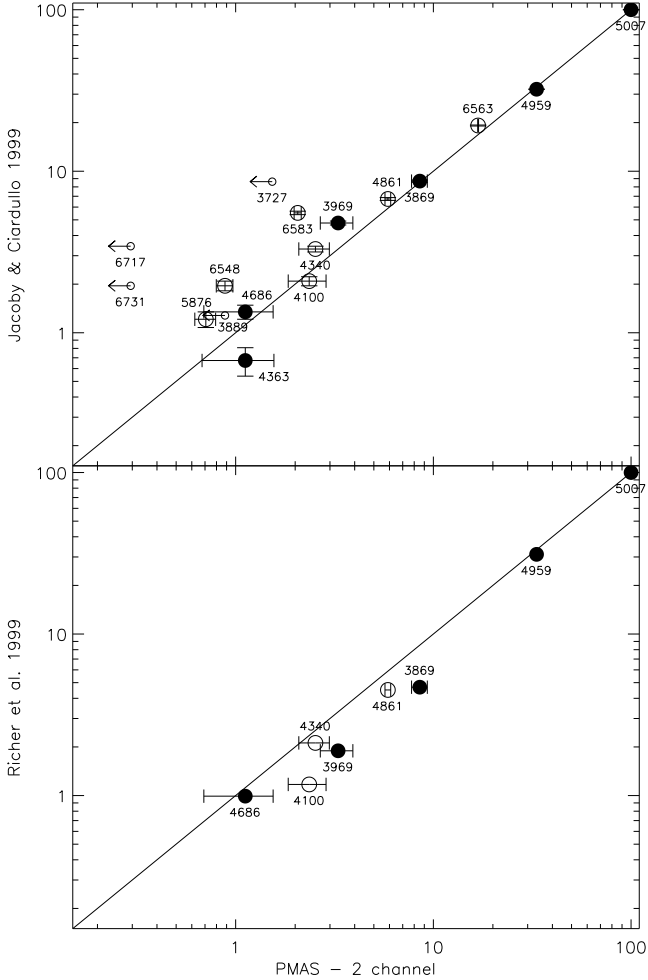}
 \figcaption{Comparison of PN29 line intensities from JC99 and RSM99 vs. PMAS 
            results, normalized to I($\lambda$5007{\AA})=100.
 \label{INTCOMP}}
 \end{figure}

In order to highlight the sensitivity to background contamination, we have
plotted the JC99 and RSM99 line intensities vs.\ our PMAS results, 
normalized to I($\lambda$5007{\AA})=100 (Fig.~\ref{INTCOMP}).
First of all, there is general agreement with JC99 for the high excitation and 
Balmer lines within the error bars, except for [Ne$\;$III]$\lambda$3967, which we measure
a factor of 0.8 fainter. Since our spectrum is somewhat noisier due to a shorter total 
exposure time (4 hours as compared to 6.5) and probably a higher blue atmospheric extinction 
at the site (Hopp \& Fernandez 2002), we cannot rule out an error. We note, however, 
that RSM99 quote a value even lower than ours (0.6$\times$ JC99), so that we do not consider 
the difference to be significant. More importantly, there is a striking disagreement
for the low excitation lines [O$\;$II]$\lambda$3727, 
[N$\;$II]$\lambda$$\lambda$6548,6583, and [S$\;$II]$\lambda$$\lambda$6717,6731,
for all of which we derive significantly lower values (or upper limits,
respectively).
We thus confirm that indeed the suspected background contamination is present
not only for [S$\;$II], but also for other ions, including [O$\;$II].


\vskip 10mm

\subsection{PN27}
\label{PN27}


 \begin{figure}[h]
 \centering\includegraphics[width=1.0\linewidth,clip]{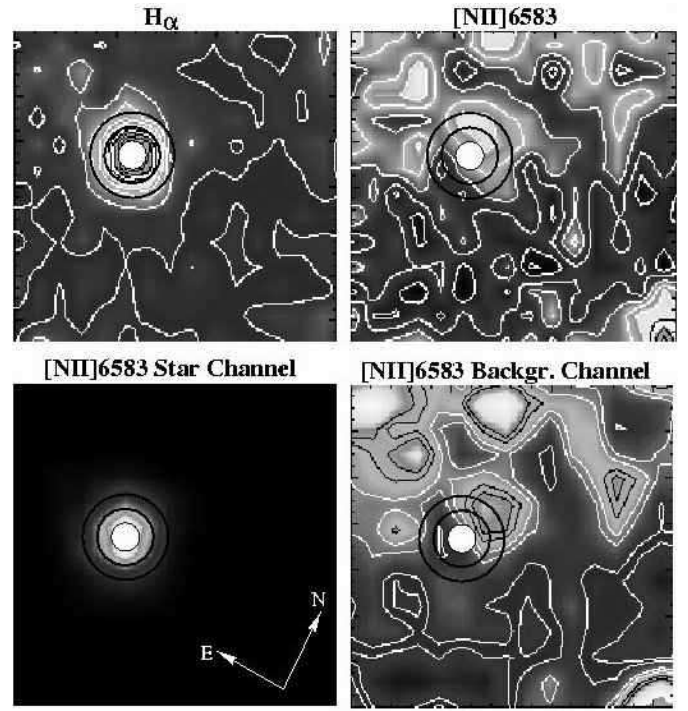}
 \figcaption{Monochromatic maps of PN27 in H$_\alpha$ and [N$\;$II] $\lambda$~6583.
 The two lower frames show how the stellar profile can be separated from the continuum
 using the {\em cplucy} 2-channel deconvolution algorithm. 
 \label{PN27MAPS}}
 \end{figure}

This object is located 152'' east of the nucleus and is among the brightest of
the CJFN89 survey (m$_{5007}$~=~20.85). JC99 were able to obtain an excellent
spectrum, allowing them to construct one of the best-constrained models of their
XPN sample. A large number of well-determined emission lines, in particular
the presence of a bright [O$\;$III]~$\lambda$4363, provided sufficient confidence
in the resulting model fit. Nevertheless, JC99 note the general faintness of
the low ionization lines like [N$\;$II] and [S$\;$II], the latter of which are
just marginally present, giving an indication of a line ratio close to the low
density limit. Our MPFS spectrum, resulting from 8~exposures with a total exposure
time of 4~hours under mediocre seeing conditions (2''-3''), is consistent 
with JC99 as far as the brighter lines are concerned (Tab.~\ref{TBL-2}).
For the fainter lines such as [S$\;$II] $\lambda$6717,6731, the S/N is not 
sufficient to draw any firm conclusions. Nevertheless, the measurable 
[N$\;$II]~$\lambda$6583 line presented us with a very good opportunity to apply 
the {\em cplucy} deconvolution technique, which uses our a priori knowledge of 
the location and shape of the PSF as a criterion for the  separation of the 
point source flux from the background.
At the wavelength of this emission line, the map shows again a background emission 
filament which is partially overlapping with the contours of the point source as 
derived from the bright nearby H$_\alpha$ line (Fig.~\ref{PN27MAPS}, upper panels). 
We applied the {\em cplucy} algorithm to investigate in how far the filamentary 
intensity is affecting a conventional measurement of the point source emission at 
the locus of PN27, and in how far it is possible to separate the two components.
The lower panels of Fig.~\ref{PN27MAPS} show separate [N$\;$II]~$\lambda$6583 maps 
for the point source and the background channel, respectively, after processing 
with {\em cplucy}. The ISM filament, which is also seen in the
direct FP image in Fig.~\ref{M31XPN_Halpha}, is very bright in proportion
to the XPN, and, owing to its vicinity to the centroid of the point source, 
practically impossible to detect with the very limited spatial resolution of a slit. 
Even our datacube aperture spectrophotometry using a background annulus resulted in 
an overestimate of the XPN emission in this line: 3.1$\times10^{-16}$erg/cm$^2$/sec.
The upper limit derived with {\em cplucy} amounts to
(1.2$\pm0.3)$$\times10^{-16}$erg/cm$^2$/sec . This result is only half the line 
strength reported by JC99 and RSM99. The important conclusion from this experiment
is, that slit spectroscopy is indeed critically subject to systematic errors in
crowded field environments, and that 3D provides a good tool to overcome these
limitations, as one would have expected from the long-standing experience with
crowded field CCD photometry. As for the particular target PN27, a more accurate 
measurement of [N$\;$II]~$\lambda$6583 and other faint emission lines would 
clearly benefit from an 8-10m class telescope with superior image quality,
e.g.\ the GMOS-IFU at the Gemini-North telescope (Allington-Smith et al.\ 2000).

\subsection{PN56}
\label{PN56}
PN56 is another test case with suspected low-ionisation background contamination, but
bright enough in [O$\;$III] to provide a well-defined PSF ( m$_{5007}$~=~20.94 ).
The XPN is located 120~arcsec south-west of the nucleus. The suspicious filament 
extends in an arc-like distribution to the West/South-West, also affecting another
nearby object (PN181) which is observed simulataneously in the
same FOV (Fig.~\ref{M31XPN_Halpha}). A third surface brightness maximum in
H$_\alpha$ is seen about 9~arcsec SW of PN56. The somewhat irregular outline of
the map in Fig.~\ref{M31XPN_Halpha} is the result of dithering, allowing us to cover 
a larger FOV than provided by the IFU alone. The observations were performed with
PMAS during comissioning of the U600 grating. Due to poor weather conditions (clouds), 
only spectra in the blue were recorded. 
The spectrum of PN56 (Fig.~\ref{XPNSPECTRA}) exhibits a relatively
bright H$_\beta$ of I([O$\;$III]$\lambda$5007)/I($\lambda$4861)=12.8, and the lines
of H$_\gamma$, H$_\delta$, [O$\;$III] $\lambda$4363, and [Ne$\;$III] $\lambda$3869.
For other lines like He$\;$II $\lambda$4686, [O$\;$II] $\lambda$3727,3729,
[Ne$\;$III] $\lambda$3968 we were merely able to derive upper limits.

\subsection{PN181}
\label{PN181}
PN181 is located 4.5~arcsec west of PN56. It is the faintest of our objects, with
m$_{5007}$~=~21.85. It is embedded in an environment with significant filamentary
emission. Since the target was included merely in a fraction of the exposures 
contributing to the mosaic with PN56, the total integration time is only 6000~sec. 
The quality of the exposure was also affected by poor transmission of the atmosphere 
and clouds. The spectrum is dominated by photon shot noise of the background 
such that only [O$\;$III]$\lambda$$\lambda$5007,4959, H$_\beta$ and 
[Ne$\;$III]$\lambda$3968 are detected with confidence (Fig.~\ref{XPNSPECTRA}). 
At the 2~$\sigma$-level, there are hints of H$_\gamma$ and H$_\delta$, but we 
refrain from further analysis at this unsufficient level of signal-to-noise.

\subsection{PN276}
\label{PN276}
Spectra of this object were secured for the first time by JS during the 1997 MPFS run 
at the 6m BTA (Schmoll 2001). PN276 had been selected after a series of 
non-detections of other targets due to pointing problems, since it is a relatively 
bright object (m$_{5007}$~=~20.48), which was expected to be more easily recognized
on the raw CCD frame. It is also located further out in the 
disk (r$_{nucl}$=820~arcsec) and therefore less subject to background subtraction
problems. We remind the reader that despite the aforementioned 
advantage of 3D with regard to pointing accuracy, direct target acquisition using a 
standard broadband TV guiding camera is extremely difficult, if not impossible, because 
of the lack of contrast, thus demanding blind offset pointing 
(cf.\ Fig.~2 in Roth et al.\ 2002b).
PN276 was also considered interesting because of the opportunity of observing
simultaneously two objects within the same FOV (PN276, PN277 in Ford \& Jacoby 1978,
of which the latter, however, was no longer listed in CJFN89).
Already from the inspection of monochromatic maps which were produced by the MPFS online 
data reduction software, a suspicious extended, triangular shape of the object became 
obvious. Considering the spatial extension, and the spectrum with untypically bright low
excitation lines (e.g.\ [O$\;$I] $\lambda$6300), it became quickly clear that this object
could not be explained as a single XPN, or the superposition of two XPNe with small 
angular separation.  As an interesting comparative test case, PN276 was subsequently 
reobserved also with MPFS in 1998, with INTEGRAL in 1998, and with PMAS in 2002. 
The resulting flux-calibrated spectra are shown in Fig.~\ref{SNRSPECTRA}. 
While the 1998 observations with MPFS and INTEGRAL were obtained with similar spectral 
resolution and wavelength coverage, the PMAS observations were chosen to extend further 
into the blue at twice the dispersion. Although this spectrum becomes increasingly
noisy towards the atmospheric cutoff, the extraordinary strength of 
[O$\;$II] I($\lambda$3727,3729)=820 is clearly visible. The fitted line intensities for 
each observation are listed in Tab.~\ref{TBL-2}, showing good agreement between the 
different instruments.

In an attempt to understand the physical nature of this object, the XPN hypothesis
can clearly be ruled out. Firstly, the [S$\;$II] doublet $\lambda$6717,6731 line ratio 
of 1.03 implies an electron density of $\approx$400~e$^-$/cm$^3$, which is low for a
typical PN. Secondly, T$_e \approx 55000$~K, derived from the
[O$\;$III] line ratio (I($\lambda$4959) + I($\lambda$5007)) / I($\lambda$4363) is
too high. Thirdly, the spatial extension of approximately 5''$\times$4'' is
incompatibel with the typical angular size of an XPN at a distance of 770~kpc (< 0.1'').
For an H$\;$II region, again, the electron temperature is too high. 
The spectrum matches very well the general signature of supernova remnants
(SNR) as explained e.g. in Osterbrock (1989): strong [S$\;$II] $\lambda$6717,6731,
[O$\;$II] $\lambda$3727,3729, and [O$\;$I] $\lambda$6300,6364 all of which are 
indicative of shock-excitation rather than photoionization. The high value of T$_e$ 
is typical for SNR, as well as the low electron density. 
With log(I(H$_\alpha$)/I([N$\;$II]))~=~-0.22 vs.\ log(I(H$_\alpha$)/I([S$\;$II]))~=~-0.1 
in the diagnostic diagram of Sabbadin \& D'Odorico (1976), PN276 is located 
unambiguously in the SNR region, far away from  H$\;$II regions or PNe. The 
projected size of 15$\times$12~pc$^2$, confirmed with our direct CAFOS images in 
[O$\;$III]~$\lambda$5007 and H$_\alpha$ (Fig.~\ref{M31XPN_Halpha}) is also only plausible 
for a SNR. 
Finally, the coincidence with an X-ray source at the same position, detected in the 
ROSAT PSPC survey of M31 (Supper 1997) and flagged as possible SNR (ID249) provides 
further evidence that PN276 is in fact a SNR: the position of 0~42~16.93~+41~18~41.5~(1975)
as listed in CJFN89 differs by only 2.7'' of the ROSAT position of 
0~42~16.9~+41~18~38.8, precessed from 2000 to 1975, in accord with the PSPC error 
circle of 5~arcsec and the astrometric accuracy of $\approx$1~arcsec quoted in CJFN89.

\section{Summary and Conclusions}
\label{SUMMARY}

Our pilot study has demonstrated the usefulness of integral field (3D) spectroscopy 
for spectrophotometry of faint planetary nebulae in M31. We have observed 5 objects,
two of which (PN27, PN29) could be compared with previously published
data. Of the three new spectra, one reveals that the object (PN276) had been 
misclassified as an XPN. Spatial extension, characteristic properties of the spectrum
and positional coincidence with a previously detected X-ray source indicate with a
high level of confidence that the source is a SNR.

Thanks to the full 2-dimensional spatial information of the observing technique,
we make explicitely use of the a priori knowledge of the PSF, which is conveniently
determined from the bright XPN [O$\;$III] $\lambda$5007 emission line. The PSF is
used as a constraint to disentangle the point source flux from contaminating
contributions that are due to diffuse or filamentary gaseous emission in the bulge of M31.  
We find that the low excitation XPN line intensities [O$\;$II] I($\lambda$3727,3729),
[N$\;$II] $\lambda$6548,6583, and [S$\;$II]$\lambda$$\lambda$6717,6731
tend to be overestimated in previous measurements with classical spectrographs. 
We have measured the emission line intensity of the ISM component and find that
the fluxes corresponding to a typical spectrograph aperture are compatible with a
level of background contamination required to explain the systematic discrepancies
in line intensities, which have been reported in the literature. Simulated slit 
spectroscopy from a datacube for PN29 allowed us to produce arbitrary 
line ratios for I([O$\;$III] 5007{\AA})/I(H$_\beta$), suggesting an explanation for 
the discrepancy reported by JC99 and RSM99. We have employed successfully the 
{\em cplucy} deconvolution algorithm to achieve the best separation of point sources
from a highly complex background, both in terms of spectral and of spatial variation.

A fundamental conclusion is that the limited geometry of conventional slit spectroscopy 
is ill-suited to solve the problem of background subtraction of faint sources embedded 
in complex, high surface brightness crowded fields. Our results support and quantify 
the previously suspected notion that such measurements suffer significantly from 
systematic errors. 3D spectroscopy offers the unique opportunity to make full use of 
the 2-dimensional spatial and 1-dimensional spectral information of a datacube to 
construct the best possible background model and eliminate these problems. At the other 
extreme, single fiber multi-object spectroscopy in high surface brightness regions is 
obviously a very limited technique and most vulnerable to systematic errors since there 
is no practical way to provide reliable background estimates in the immediate vicinity 
of the point source (but see Durrell et al.\ 2003 for excellent results from a 
{\em radial velocity} survey based on  [O$\;$III] of XPN in M51). It must be stressed, 
however, that 3D does not necessarily provide dramatic improvements in S/N for 
photon-starved observations, but rather allows one to {\em eliminate the systematic 
errors} discussed in this paper.

As a side-remark, we note that, in contrast to the conclusions of Allington-Smith \& 
Content (1998a), residual cross-talk between adjacent spectra on the CCD is potentially 
dangerous for PSF-fitting algorithms, since the net effect is a distortion of the PSF 
(cf. Fig.~\ref{PSFANALYSIS}), which must be expected to vary with FOV, wavelength, and 
time. Analogous to the reasoning of Mateo (1998) concerned with CCD photometry, the 
distortion is not important for the fitting of faint stars, however very much so for 
applications requiring a good definition of the PSF (cf. Wisotzki et al.\ 2003).


In continuation of our successful pilot study, we are pursuing a guaranteed observing 
time programme targeting XPN in M31 and other local group galaxies. The analysis
of our present and the expected new data by comparison with ionization models following 
the receipe described by JC99 will be published in a forthcoming paper.

We have begun to use 3D spectrophotometry for a number of other applications which
benefit from the evaluation of spatial information and the PSF, e.g.\ spectrophotometry
of SN2002er (Christensen et al.\ 2003; see also Aldering et al.\ 2003 for a description
of the SNAP experiment), the quadruple lensed QSO HE~0435-1223 (Wisotzki et al.\ 2003), 
optical counterparts of superluminous X-ray sources (Lehmann et al., in prep.), 
LBV candidates in M33 (Fabrika et al., in prep.). 
Although our XPN study was facilitated by the availability of the bright emission 
in [O$\;$III] $\lambda$5007{\AA}, which is a rather fortunate case to determine 
a well-defined PSF, we stress that the PSF fitting technique, in particular when using 
the two-channel {\em cplucy} algorithm in combination with high-resolution HST images,
opens a new window for solving the difficult observational problems of resolved stellar 
populations in nearby galaxies.

Although an upgrade of the present IFU with a 32$\times$32 element lens array is
underway, increasing the diameter of the FOV by a factor of 2, the multiplex 
advantage of PMAS for the rather sparsely distributed XPNe in nearby 
galaxies is very limited. This fact is a direct consequence of the baseline design, 
which favours wavelength coverage over the maximum number of spectra in terms of 
detector space, making PMAS a rather unique instrument. Normally, PMAS can 
observe only one target at a time. With a typical rate of one object per night, 
the efficiency may appear prohibitively low. 
On the other hand, we note that in a recent paper by Magrini et al.\ (2003), 
out of 42 XPN candidates in M33, observed during 2 nights with the WYFFOS multi-object 
fiber spectrograph at WHT, only 3 spectra of were found to be of sufficient quality for 
an abundance analysis.

Clearly, the spectrophotometry of our XPN targets would benefit substantially from
the light-collecting power and image quality of a modern 8-10m class telescope, 
pushing the source confusion problem to fainter limits and providing better S/N
for the weak lines which are so crucial for the plasmadiagnostic analysis 
([O$\;$III]$\lambda$4363, [S$\;$II] $\lambda$6717,6731, etc.). 
Multi-object spectroscopy of supergiant stars has already been demonstrated 
to be feasible out to distances beyond the local group with reasonable
integration times on the order of $\approx$10~hours (Bresolin et al.\ 2001). 
The single-target approach of our pilot study would be unrealistically expensive. 
However, the use of a multi-object deployable IFU instrument in combination with 
the techniques described in this paper would eventually become an extremely powerful 
tool for crowded-field spectroscopy and the study of resolved stellar populations in 
local galaxies, one of the key science cases of the proposed new generation of 
Extremely Large Telescopes with diameters of 30-100m. As a promising alternative strategy, 
complex 3D instruments with arcmin FOV and of order 10$^5$-10$^6$ spatial elements
are currently under investigation for the 2nd Generation Instrumentation Plan 
at the ESO-VLT (Bacon et al.\ 2002, Morris et al.\ 2002).

\acknowledgments
      Part of this work was supported by the German
      \emph{Deut\-sche For\-schungs\-ge\-mein\-schaft, DFG\/} under
      grant HA1850/10-3 , and by the German Ver\-bund\-for\-schung 
      des BMBF, grant 05AL9BA1.
      The authors would like to thank Till Soffner, University of Munich, 
      and Hans Hippelein, MPIA Heidelberg, for providing us with the H$_\alpha$
      Fabry-P\'erot image of M31 .
      MMR, TB and JS are indebted to Victor Afanasiev and Seguei 
      Dodonov, Special Astrophysical Observatory in Selentchuk, Russia,
      for their hospitality and help with MPFS observations at the 6m BTA, 
      for an introduction into 3D data reduction, and for fruitful discussions 
      and insight into integral field spectroscopy. We are thankful to 
      Evencio Mediavilla for generously providing us with INTEGRAL observing time.
      JS would like to thank Evencio Mediavilla and Stefano Ciroi for introducing
      him into IFU data reduction techniques under IRAF.
      The excellent support of Calar Alto staff during the PMAS commissing 
      phase and in operation is gratefully acknowledged.

\clearpage
\onecolumn


\end{document}